\documentclass[11pt,a4paper]{article}
\pdfoutput=1
\usepackage{jheppub}
\usepackage{amsfonts}
\usepackage{bbm}
\usepackage{graphicx}
\usepackage{caption}
\usepackage{subcaption}
\usepackage{bigints}
\usepackage[normalem]{ulem}

\def\a{\alpha}
\def\b{\beta}

\def\v{\varphi}

\def\l{\lambda}    
\def\m{\mu}

\def\o{\omega}  
  
\def\th{\theta}                  
                                     
\def\s{\sigma}                                  
\def\t{\tau}

\def\vth{\vartheta}

\newcommand{\dd}{{\mathrm d}}

\newcommand{\CN}{\mathcal{N}}

\newcommand{\CH}{\mathcal{H}}

\renewcommand{\Im}{{\rm Im}}
\renewcommand{\Re}{{\rm Re}}
\newcommand{\Tr}{\mbox{Tr}}

\def\CP{\mathbb{CP}}

\newcommand{\IR}{\mathbb{R}}
\newcommand{\IC}{\mathbb{C}}
\newcommand{\IZ}{\mathbb{Z}}
\newcommand{\IH}{\mathbb{H}}

\newcommand{\IP}{\mathbb{P}}

\newcommand{\half}{\frac{1}{2}}
\newcommand{\ndt}{\noindent}

\newcommand{\nn}{\nonumber}

\def\p{\partial}

\def\bea{\begin{eqnarray}}
\def\eea{\end{eqnarray}}
\def\be{\begin{equation}}
\def\ee{\end{equation}}
\def\ba{\begin{align}}
\def\ea{\end{align}}

\newcommand{\bem}{\begin{pmatrix}}
\newcommand{\eem}{\end{pmatrix}}

\def\qb {\bar{q}}

\def\={\;  = \;}
\def\+{\, + \,}

\def\wt{\widetilde}

\def\bar{\overline}

\def\rt2{\sqrt{2}}

\renewcommand{\Im}{\mbox{Im}}
\renewcommand{\Re}{\mbox{Re}}

\newcommand{\bPhi}{\overline{\Phi}}
\newcommand{\bP}{\overline{P}}

\newcommand{\bSigma}{\overline{\Sigma}}

\def\inn{\,\in\,}
\def\mypmod#1{\; (\mod \; {#1})}

\title{Squashed toric sigma models and mock modular forms}

\author{Rajesh Kumar Gupta and Sameer Murthy}
\affiliation{Department of Mathematics, King's College London,
The Strand, London WC2R 2LS, UK}

\emailAdd{rajesh.gupta, sameer.murthy at kcl.ac.uk}

\abstract{
We study a class of two-dimensional~$\CN=(2,2)$ sigma models 
called squashed toric sigma models, using their Gauged Linear Sigma Models (GLSM) description. 
These models are obtained by gauging the global~$U(1)$ symmetries of toric GLSMs and introducing a set 
of corresponding compensator superfields. The geometry of the resulting vacuum manifold is a 
deformation of the corresponding toric manifold in which the torus fibration maintains a constant size 
in the interior of the manifold,  thus producing a neck-like region.
We compute the elliptic genus of these models, using localization, in the case when the unsquashed vacuum 
manifolds obey the Calabi-Yau condition. The elliptic genera have a non-holomorphic dependence 
on the modular parameter~$\tau$ coming from the continuum produced by the neck. 
In the simplest case corresponding to squashed~$\IC/\IZ_{2}$ the elliptic genus is 
a mixed mock Jacobi form which coincides with the elliptic genus of the~$\CN=(2,2)$~$SL(2,\IR)/U(1)$ 
cigar coset. 
}

\begin{document}

\maketitle

\section{Introduction and summary \label{sec:intro}}

The classic work of~\cite{Schellekens:1986yi,Witten:1986bf,Alvarez:1987wg}  
on the elliptic genus led to the establishment of a three-way relation between 
two-dimensional~$\CN=(2,2)$ superconformal field theories (SCFTs), compact Calabi-Yau (CY) manifolds, 
and modular and Jacobi forms. 
The elliptic genus of an~$\CN=(2,2)$ SCFT~$M$ with central charge~$c$ is defined as 
\be \label{ellgendef}
\chi_\text{ell}(M;\tau,z) \; := \; \Tr_{\CH_{RR}} \, (-1)^{F} \,  q^{L_0} \,  \qb^{\overline L_0} \, \zeta^{J_0} \, , 
\qquad  q=e^{2 \pi i \tau}, \quad \zeta = e^{2 \pi i z} \, ,
\ee
where~$\CH_{RR}$ is the Ramond-Ramond Hilbert space of the theory,~$L_{0}$ and~${\overline L_{0}}$ 
are the left and right-moving Hamiltonians of the~$(2,2)$ algebra,~$J_{0}$ is the left-moving R-charge, 
and~$F$ is the fermion number operator. 
The elliptic genus of a two-dimensional $(2,2)$ SCFT in the moduli space of a compact 
CY manifold of complex dimension~$d$ is a Jacobi form of weight~0 and index~$d/2=c/6$.

In this paper we focus on a class of~$\CN=(2,2)$ supersymmetric sigma models 
which flow to SCFTs with \emph{non-compact} target space. 
Non-trivial non-compact target spaces appear in diverse physical situations, e.g.~as the vacuum 
manifold of supersymmetric quantum field theories, as supersymmetric black hole moduli 
spaces, and  as backgrounds on which strings and D-branes can propagate (some examples 
are the near-horizon regions of NS5-branes, ALE spaces, and the conifold).  
The presence of a continuum of operators in the spectrum, which is a signature of the non-compactness, is a 
source of many subtleties 
and often invalidates basic conclusions that apply to theories with compact target space. 
An example of this phenomenon that we will study in particular in this paper is the holomorphicity of the elliptic genus.

In theories with a discrete spectrum, holomorphicity (in~$\tau$) of the elliptic genus follows from the simple 
argument~\cite{Witten:1986bf} that all states with non-zero~$\overline L_0$ come in representations with 
equal number of bosonic and fermionic states, and 
therefore do not contribute to the elliptic genus. When there is a continuum component in the spectrum, 
the trace in Equation~\eqref{ellgendef} needs a precise definition, and the measure involved in the integral over 
the continuum may not be equal for bosonic and fermionic states with equal values of~$\overline L_0$. (Physically
this means that the density of states of fermions and bosons are not equal, this happens when there is 
a non-zero phase shift in scattering from infinity.) This can lead to an incomplete cancellation, and therefore 
to a~$\overline \tau$-dependence of the elliptic genus. 

This phenomenon has been understood in great detail for the supersymmetric version of the Euclidean 2d black hole, 
also known as the \emph{cigar}, whose target space is the (Euclidean) $SL_{2}(\IR)/U(1)$ $\CN=(2,2)$ SCFT. 
The elliptic genus of the cigar theory was calculated in \cite{Troost:2010ud,Eguchi:2010cb,Ashok:2011cy} by 
computing the relevant functional integral of the WZW coset, using the technique of~\cite{Gawedzki:1991yu}. 
The same phenomenon was also found, in a spacetime avatar, in string theories in the near-horizon 
geometry of NS5-branes~\cite{Harvey:2013mda,Harvey:2014cva}  and their T-duals~\cite{Cheng:2014zpa},
all of which involve the cigar SCFT as an important component. 
The elliptic genus of the cigar coset, as well as some generalizations, was later computed in a much 
simpler manner using the GLSM description~\cite{Murthy:2013mya, Ashok:2013pya}. 
The class of modular objects that captures the modular but non-holomorphic behavior 
of the elliptic genus of the cigar theory was discovered relatively recently, and is called mock modular 
forms~\cite{Zwegers:2002, Zagier:2007} (more precisely the cigar elliptic genus is a mixed mock 
Jacobi form~\cite{Dabholkar:2012nd}). 
These functions transform like holomorphic modular forms, but 
their~$\overline{\tau}$-derivative is non-vanishing and can be summarized by a 
holomorphic anomaly equation (summarized in Appendix~\ref{sec:Jacobi}). 
In this sense the study of the cigar theory has led to an extension of the three-way relation mentioned 
in the beginning. 

While it is nice to see mock modular forms fit into a corner of conformal field theory 
and string theory, it is worth noting that one of the links in this refinement of the three-way relation has not 
been well-understood, namely the role of geometry.  
This paper aims to reduce this gap by studying a class of non-compact manifolds which are thought to flow to 
SCFTs with a non-trivial dilaton profile, and whose elliptic genus shows interesting new 
modular behavior. 
We find functions which depend explicitly on~$\bar{\tau}$ but transform like a holomorphic 
multi-variable Jacobi form. The~$\bar{\tau}$-dependence is captured by a differential equation 
(Equation~\ref{holanom}) that generalizes the one obeyed by mock Jacobi forms.
We hope that our physical ideas and results serve as a motivation for further work that is needed to understand the 
links between non-compact manifolds and SCFTs, their elliptic genera, and mock modular forms and 
their generalizations. These considerations may also be useful for the Umbral  
Moonshine program~\cite{Eguchi:2010ej, Cheng:2012tq, Cheng:2013wca} where one is searching for 
geometric or physical objects that give rise to particular mock Jacobi forms.

\vspace{0.2cm}

\begin{figure}[h]
\centering
\includegraphics[width=8cm]{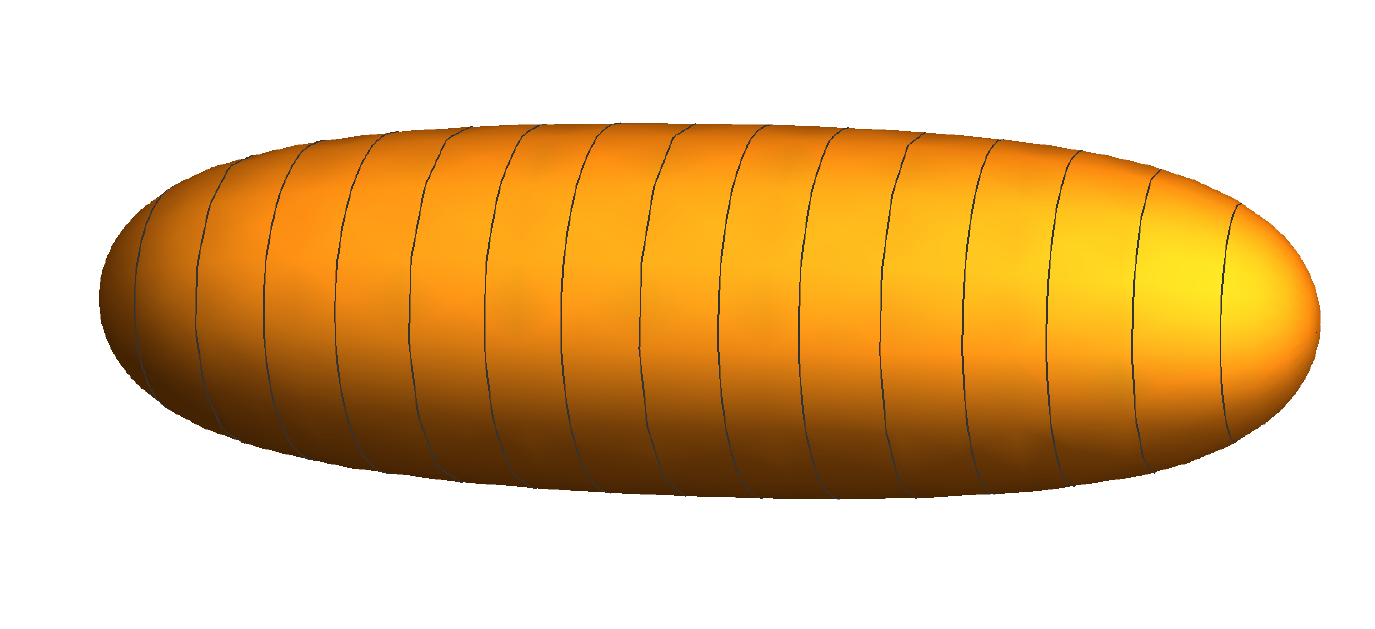}
\caption{Squashed $\IC\IP^{1}$ with $\widehat k=1/4$ and $r=1$}
\label{fig:Sausage}
\end{figure}

The objects of study in this paper, called squashed toric manifolds, are deformations of toric manifolds. 
While toric manifolds have been
studied extensively by geometers as well as by string theorists, their squashed counterparts have received less 
attention. The defining feature of a (real) $2d$-dimensional toric manifold is a non-trivial~$U(1)^{d}$ 
action at a generic point. The geometric structure is that of a~$d$-dimensional 
torus fibered on a~$d$-dimensional base. This torus varies in size as we move along the base, and 
there are distinguished fixed points where the torus shrinks to zero size. 
The squashing deformation makes the torus gain a constant size in the deep interior of the manifold. A 
simple illustrative example is that of squashed~$\IC \IP^{1}$ in which the spherical shape gets 
squashed to a sausage (see Fig \ref{fig:Sausage}). 

The GLSMs for these squashed toric manifolds were introduced in \cite{Hori:2001ax}. 
The starting (unsquashed) point is a model with~$n$ chiral superfields and~$n-d$ gauge superfields, 
whose vacuum manifold is~$2d$-dimensional. The squashing deformation gauges the~$d$-dimensional 
flavor symmetry of this theory and at the same time introduces a set of~$d$~compensator-chiral superfields 
(which translate under the flavor gauge fields). The vacuum manifold of the squashed model thus 
remains~$2d$-dimensional, and is called the squashed toric manifold. 
In most of the paper we study manifolds that obey the 
Calabi-Yau condition, namely the vanishing of the sum of the gauge charges of the chiral multiplets for each 
gauge field. This condition ensures that the 2d QFTs has non-anomalous chiral~$U(1)$ R-symmetries 
so that they can flow to~$\CN=(2,2)$ SCFTs. 
This sum-rule in turn implies that the initial toric manifold (and therefore its squashed deformation) 
is non-compact~\cite{Morrison:1994fr}. 
The squashing deformation is a very strong deformation in that it reaches the asymptotic region, for example, if 
we begin with a space of the form~$\IC/\IZ_{2}$ the squashed counterpart has a cylinder-like shape
asymptotically. This is in a different universality class of theories compared to the toric CY 
manifolds---the Ricci curvature is no longer zero and is supported by a non-trivial dilaton profile. 
Our main result, that we now describe briefly, is an expression for the elliptic genus of these squashed toric 
manifolds. 

First we recall some results for the elliptic genus of toric CY manifolds.  As mentioned above, these
manifolds are necessarily non-compact. Although one can formally write down the elliptic genus as in 
Equation~\eqref{ellgendef} using the~$\CN=(2,2)$ superconformal algebra, the quantity is usually  
ill-defined because of the infinite volume of the space\footnote{The presence of the geometric singularity in the 
case of orbifolds may also be seen as a potential problem in geometry. Here we take the attitude that such 
singularities can be removed by usual CFT methods, either by turning on an FI term deformation or by resolving 
the space. One still has, however, the issue of the bosonic zero-modes which make the elliptic genus formally
infinite.}. 
One way to regulate this divergence is to turn on a background Wilson line~$u$ of the external gauge field 
which couples to the angular momentum in the spacetime (which is a flavor charge~$F$ in the sigma model). 
The zero modes of the bosons are charged under this symmetry, and therefore the divergence is lifted. 
The modified elliptic genus
\be \label{ellgentorEq}
\chi_\text{ell}(M_\text{tor};\t,z,u) \= 
\Tr_{\CH_{RR}} \, (-1)^{F} \,  q^{L_0} \, \qb^{\overline L_0} \, \zeta^{J_0} e^{2 \pi i u F}
\ee
is now a sensible supersymmetric index. 
This route was used to study the elliptic genus of ALE and ALF spaces in \cite{Harvey:2014nha}, and 
the answer thus obtained is a Jacobi form meromorphic in the elliptic variable~$u$. 
This meromorphicity is related to the fact that the scale introduced by the background Wilson line~$u$ 
mildly breaks the infinite-dimensional superconformal algebra, as can be seen from the fact that bosons 
can no longer be separated into left- and right-moving parts~\cite{Dabholkar:2010rm,Haghighat:2015ega}.

In this paper we take a different route. The squashing deformation is now an operation intrinsic to the theory
and therefore one does not need to introduce any external scale to regulate the elliptic genus. 
The original definition~\eqref{ellgendef} of the elliptic genus applies to the squashed models as long as 
we define the trace carefully. The answer turns out to be holomorphic in the elliptic variable~$z$, but 
non-holomorphic in the modular parameter~$\t$.

To be concrete, we consider a~$2d$-dimensional toric GLSM~$M_\text{tor}$, and associate the set of chemical 
potentials~$\{u'_{\ell}\}$, $\ell = 1, \cdots d$, 
to the corresponding~$U(1)$ symmetries. 
This GLSM has a flavored elliptic genus~$\chi_\text{ell}(M_\text{tor};\tau,z,\{u'_{\ell}\})$ as described above. 
The~$U(1)$ toric symmetries act on the fields of the GLSM as 
global flavor symmetries with the chiral superfields having charges~$F_{i}^{\ell}$, $i=1,\cdots n$. 
Two sets of parameters that are particularly important for the elliptic genus of the squashed 
manifold~$\wt M_\text{tor}$ are the sum of the charges~$b_\ell=\sum_{i=1}^nF_i^\ell$ for 
each flavor symmetry, and the strength of the coupling~$\{k_{\ell}\}$ of the compensator fields. 
We define~$\wt k_{\ell} = k_{\ell}/b_{\ell}^{2}$ which is the effective strength of the squashing 
deformation, and which determines the size of the constant circles in the squashed manifolds. 
Our main result is an expression for the elliptic genus of~$\wt M_\text{tor}$ given in terms of 
an integral over the~$d$-dimensional torus~$E_\tau^{d}$, with~$E_\tau=\IC/(\IZ\t+\IZ)$, spanned 
by the holonomies~$\{u'_{\ell}\}$ of the flavor symmetry gauge fields. 
Define, for~$z,u \in \IC$, the non-holomorphic kernel function:
\be \label{defHell}
H_k (\tau, z, u) \= k \sum_{m,w \, \in \, \mathbb{Z}} \,
e^{2\pi i w z - \frac{\pi k}{\tau_2} \bigl( w\tau+m+u+\frac{z}{k} \bigr)
\bigl(w \bar{\tau}+m+\overline{u}+\frac{z}{k} \bigr)} \, .
\ee 
The elliptic genus of the squashed toric model~$\wt M_\text{tor}$, given in Equation~\eqref{SquashElliptic3} 
of the text, is a~$d$-dimensional convolution of the elliptic genus of~$M_\text{tor}$ with these kernel functions:
\bea
\chi_\text{ell}(\wt M_\text{tor}; \tau,z)  \= 
\bigintsss_{E_\tau^{d}} \, \prod_{\ell=1}^d   \frac{\dd^2 u'_\ell}{\t_{2}} \,  
H_{\wt k_\ell} (\tau, z, u'_\ell)  \; \chi_\text{ell}(M_\text{tor};\tau,z, \{u'_{\ell}/b_{\ell} \}) \, . 
\eea
This convolution thus shifts the signature of the non-compactness from a meromorphicity in~$u$ 
to a non-holomorphy in~$\t$.

Apart from their intrinsic mathematical interest, these results naturally prompt the conjecture for the existence 
of an RG flow from 
the squashed toric sigma models to certain SCFTs with the above elliptic 
genera. In the simplest example, the RG flow is from a squashed~$\IC/\IZ_{2}$ to the cigar coset SCFT. 
This RG flow is similar to that discussed in \cite{Hori:2001ax} but our UV starting point~\eqref{squashedC2Z2} 
is slightly different. 
One can make similar conjectures for each of the GLSMs of RG flows from higher-dimensional squashed 
toric manifolds to SCFTs in the IR. In~\cite{Gromov:2017} we study, and provide further evidence for, such RG flows.

\vspace{0.2cm}

The plan of the paper is as follows. In Section~\ref{sec:GLSM} we review toric manifolds, some of their properties, 
and their construction as vacuum manifolds of GLSMs. 
We then review the squashing deformation of~\cite{Hori:2001ax}. 
In each case we illustrate the discussion with four examples.
In Section~\ref{sec:EllGenTor} we present the computation of the elliptic genus of toric CY manifolds 
using the GLSM construction and the technique of localization. 
In Section~\ref{sec:EllGen} we compute the elliptic genus of the squashed toric models. 
In Section~\ref{sec:Compact} we analyze a compact (non-Calabi-Yau) example, namely the supersymmetric 
sausage model, and compute its Witten index. 
In four appendices we briefly review some details of (multi-variable) Jacobi forms and mock Jacobi forms, 
the holomorphic construction of toric manifolds,
the metric of squashed manifolds, and the action of the GLSMs,  that are referred to in various places in 
the main text.

\section{A class of gauged linear sigma models \label{sec:GLSM}}

In this section we review some basic facts about toric manifolds and toric sigma models. 
This is a wide and deep subject and we refer the reader to \cite{Morrison:1994fr,Hori:2003ic} for 
an introduction to this topic for physicists, and to~\cite{daSilva:2001} for a mathematical treatment 
of the symplectic viewpoint on toric manifolds which we follow.
Here we shall briefly review the GLSM construction of toric manifolds and
present four examples ($\IC\IP^{1}$, $\IC/\IZ_{2}$, the~$A_{1}$ space~$\IC^{2}/\IZ_{2}$, 
and the conifold) which we will use in the rest of the paper to illustrate our general results.

\subsection{A brief review of toric manifolds and toric sigma models \label{reviewtor}}

Consider the~$\CN=(2,2)$ GLSM with field content consisting of the 
chiral superfields $\Phi_{i}$, $i=1,\cdots,n$, and the abelian vector 
superfields~$V_{a}$, $a=1,\cdots,n-d$ 
with associated field strength twisted chiral superfields~$\Sigma_{a}$. 
The chiral superfields have charges~$Q^{i}_{a}$ under the vector superfields.
The action is:
\be \label{action0}
S_{0} \= \frac{1}{2\pi} \int\dd^2x\, \Biggl[ \int \dd^4\theta \, \biggl( \sum_{i=1}^{n} \bPhi_i \, 
  \exp \Bigl(\sum_{a=1}^{n-d}Q_{i}^{a} V_{a} \Bigr) \Phi_i
 \, - \, {1\over 2e^2}\sum_{a=1}^{n-d} \bSigma_{a} \, \Sigma_{a} \biggr)  \+ \half \int\dd^2\wt\theta \sum_{a=1}^{n-d} t_{a} \Sigma_{a}  \+ \text{c.c.} \Biggr] \, ,
\ee
with~$t_{a}=r_{a}-i \vth_{a}$. Here $r_{a}$ is the Fayet-Ilopoulos parameter and 
$\vth_{a}$ is the theta angle for the gauge field~$V_{a}$.

The manifold of inequivalent vacua (the vacuum manifold) is obtained 
by solving the constraints imposed by setting the 
D-terms\footnote{Throughout this paper we assume that the values of~$r_{a}$ are 
such that all the scalars~$\s_{a}$ in the vector multiplet are zero at the mimima of the potential, i.e.~there is 
no Coulomb branch.}$^{,}$\footnote{We impose Wess-Zumino gauge throughout the paper, so that the only auxiliary
fields that remains in each vector multiplet is the~$D$ field.} :
\be
D_{a} \= - e_{a}^{2} \, \mu_{a} \, , \qquad  \mu_{a} \= \sum_{i=1}^{n} Q_{i}^{a} \, |\phi_{i}|^{2} - r_{a} \, , \qquad a = 1, \dots, n-d \, , 
\ee
to zero, and quotienting by the gauge group~$G=U(1)^{n-d}$. 
Denoting the vector of FI terms with components~$r_{a}$ by~$r$, we write 
the~$2d$ real-dimensional vacuum manifold~$V$ as:
\be
V(r) \= \mu^{-1}(0)/G \, . 
\ee
It is a non-trivial fact that this vacuum manifold has a natural symplectic structure induced from that 
of~$\IC^{n}$ on which the 
chiral fields~$\phi_{i}$ live. It inherits a non-trivial~$U(1)^{d}$ Hamiltonian action from 
the~$U(1)^{d}$ flavor symmetry of the GLSM~\eqref{action0}. 
Such a manifold~$V$ is called a \emph{symplectic toric manifold}, and  
the above construction of~$V$ is precisely what is called the \emph{symplectic quotient construction} of 
toric manifolds\footnote{There is an independent algebro-geometric construction of such toric manifolds,  
called the \emph{holomorphic construction} which we briefly recall in Appendix~\ref{holtor}. It is a non-trivial fact 
that these two constructions are equivalent under certain conditions~\cite{Morrison:1994fr}. We shall not 
discuss this  in this paper.} \cite{daSilva:2001}.

As~$\IC^{n}$ is also a K\"ahler manifold and the~$U(1)^{n-d}$ action preserves its complex structure, 
the quotient space~$V$ is also a K\"ahler manifold. 
The metric can thus be written in terms of local complex coordinates~$Z^{I}(\phi_{i})$,
$I=1,\cdots d$, as a derivative of the K\"ahler potential~$K(Z_{I},\overline{Z_{I}})$:
\be
g_{I\overline{J}} \= \p_{I} \p_{\overline{J}} \, K (Z_{I},\overline{Z_{I}}) \,.
\ee
This will be the case at a generic point in field space for all the models that we discuss in this paper. 
(The manifolds we discuss typically will also have special points where there are orbifold singularities.)
The metric on the quotient space can thus be computed by implementing the quotient construction on~$\IC^{n}$, 
or equivalently, by starting with the action~\eqref{action0} and integrating out 
the gauge fields.  We will illustrate both the methods in the following examples.

\vspace{0.2cm}

\ndt {\bf Example 1:} ${\bf \CP^{1}}$, $(n,d)=(2,1)$. 

Our first example is~$\IC \IP^{1}$ which can be modelled by one vector superfield
and two chiral superfields both with gauge charges~$+1$. The two-dimensional vacuum manifold is
\be
 V(r)\=\{|\phi_1|^2+|\phi_2|^2-r=0\}/U(1)\,.
\ee
In order to implement the symplectic quotient construction, we 
write~$\phi^{i}=\rho_{i}e^{i\theta_{i}}$ with the radial variables~$\rho_{i} \ge 0$ and the angular 
variables~$\theta_{i} \in [0,2\pi]$. The symplectic form on the original~$\IC^{2}$ is: 
\be
\o \= \sum_{i=1}^{2} \,  \rho_{i} \, d\rho_{i} \wedge d\theta_{i}  \, . 
\ee
The~$D$-term constraint~$\rho_{1}^{2} + \rho_{2}^{2}=r$ implies~$\rho_{1} d\rho_{1} + \rho_{2} d\rho_{2}=0$. 
The induced symplectic form is given by
\be
\o \= \rho_{1} \, d\rho_{1} \wedge d(\th_{1} -\th_{2} )\, . 
\ee
We now express this in terms of the gauge invariant holomorphic variable
\be
Z\=\phi_{1}/\phi_{2} \, \equiv R \, e^{i \psi} \, .
\ee
Using the D-term constraint we obtain
\be \label{Rrho1rel}
R \= \frac{\rho_{1}}{\sqrt{r-\rho_{1}^{2}}} \quad \Longrightarrow \quad \rho_{1} \= \frac{\sqrt{r} \, R}{\sqrt{1+R^{2}}} \, .
\ee
The symplectic form can now be written as:
\be
\o \= \frac{ir}{2} \, \frac{dZ \wedge d\overline{Z}}{(1+Z \overline{Z})^{2}} \, ,
\ee
which is identified as the Fubini-Study  form on~$\IC\IP^{1}$, derived from the K\"ahler potential:
\be
K=r\log (1+Z\overline{Z}) \, .
\ee
The corresponding metric is :
\be
ds^{2} \ = r \, \frac{dR^{2} + R^{2} \, d\psi^{2}}{(1+R^{2})^{2}} \,.
\ee
\begin{figure}[h]
\label{figcp1}
\centering
\includegraphics[width=4cm]{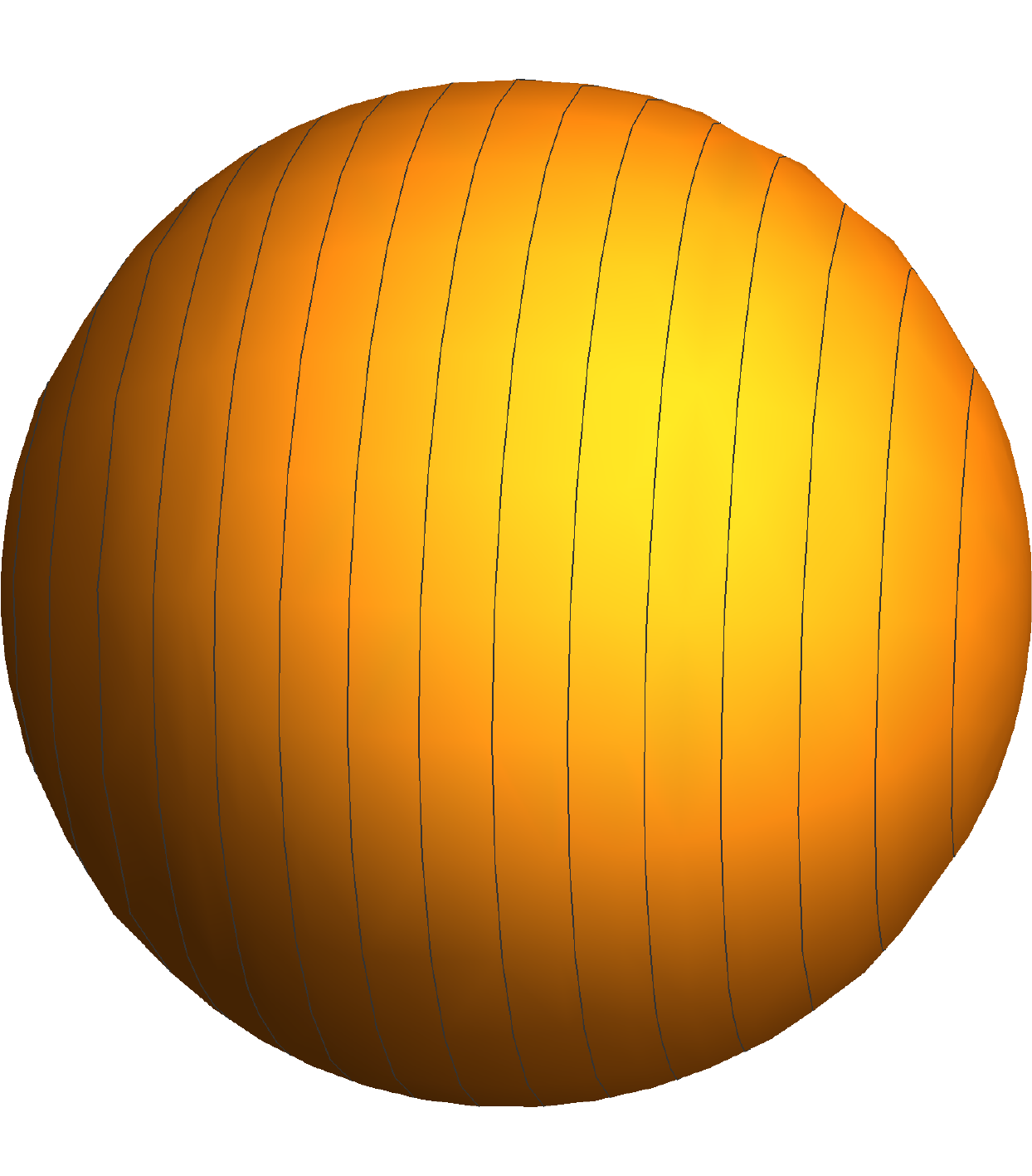}
\caption{$\IC\IP^{1}$ with~$r=1$}
\end{figure}

We note that the analysis above was classical, and the FI parameter runs in the quantum theory,
and the~$\IC\IP^{1}$ theory actually flows to a massive theory in the infra-red. 
In this paper we will mostly be interested in GLSMs that flow to 2d SCFTs in the infra-red (although 
we will make some comments on the~$\IC\IP^{1}$ model in Section~\ref{sec:Compact}). 
Now, a necessary condition for the GLSM~\eqref{action0} to flow to a 2d SCFT (or equivalently, for the 
corresponding toric variety to be a Calabi-Yau manifold) is that there is a left- and right-moving
chiral R-symmetry. For these chiral symmetries to be non-anomalous one has the condition:
\be \label{sumrule}
\sum_{i} Q^{a}_{i} \= 0 \, , \qquad a = 1, \cdots, n-d \,.
\ee
As mentioned in the introduction, such a constraint on the charges cannot be satisfied by a compact toric 
variety.
Our next three examples, as well as our main focus in the bulk of this paper, 
involve GLSMs described by the action~\eqref{action0} with the condition~\eqref{sumrule}, which describe
non-compact target spaces. 

\vspace{0.2cm}

\ndt {\bf Example 2:} ${\bf \IC/\IZ_{2}}$, $(n,d)=(2,1)$.

Our second example has two chiral superfields $\Phi_1$ and $\Phi_2$ with 
charges~$Q_i = \pm 1$ respectively. The vacuum manifold is:
\be
\quad V(r)=\{|\phi_1|^2-|\phi_2|^2=r\}/U(1)\,.
\ee 
This is one dimensional complex space with a natural gauge invariant complex coordinate $M=\phi_1\phi_2$.
We can obtain the (tree level) metric on~$V$ by the symplectic quotient as before or, equivalently, 
by integrating out the gauge fields as we now show. At a generic point on the Higgs branch~$M$ has 
a non-zero expectation value and the~$U(1)$ gauge field~$V_{\mu}$ gets a mass of order~$e$. At lower energy scales, 
we can integrate out the gauge field by solving the classical equations of motion to obtain:
\be
V_\mu \=
-\frac{1}{2} \, \frac{\sum_{i=1}^2 Q_i \bigl( \bar\phi_i\p_\mu\phi_i-\phi_i\p_\mu\bar\phi_i \bigr)}{\sum_{i=1}^2 Q_i^2|\phi_i|^2} \, .
\ee
Solving the D-term equation for $\phi_1$ and $\phi_2$ in terms of FI parameter~$r$ and~$M$, we obtain:
\be
|\phi_1|^{2} \= \frac{r}{2}+\sqrt{|M|^2+\frac{r^2}{4}}, \qquad |\phi_2|^{2} \=  -\frac{r}{2}+\sqrt{|M|^2+\frac{r^2}{4}}
\ee
Substituting this in the action we get the non-linear sigma model with the target space metric 
\be\label{metricC2Z2r}
ds^2\=\frac{dM\,d\bar M}{\sqrt{4M\,\bar M+r^2}}\,,
\ee
which can be derived from the K\"ahler potential
\be
K \= \sqrt{4 M\,\bar M+r^{2}} -  |r| \, {\rm arctanh} \Big( \frac{\sqrt{4 M\,\bar M+r^{2}}}{|r|} \Bigr)  \, . 
\ee
\begin{figure}[h]
\centering
\includegraphics[width=5cm]{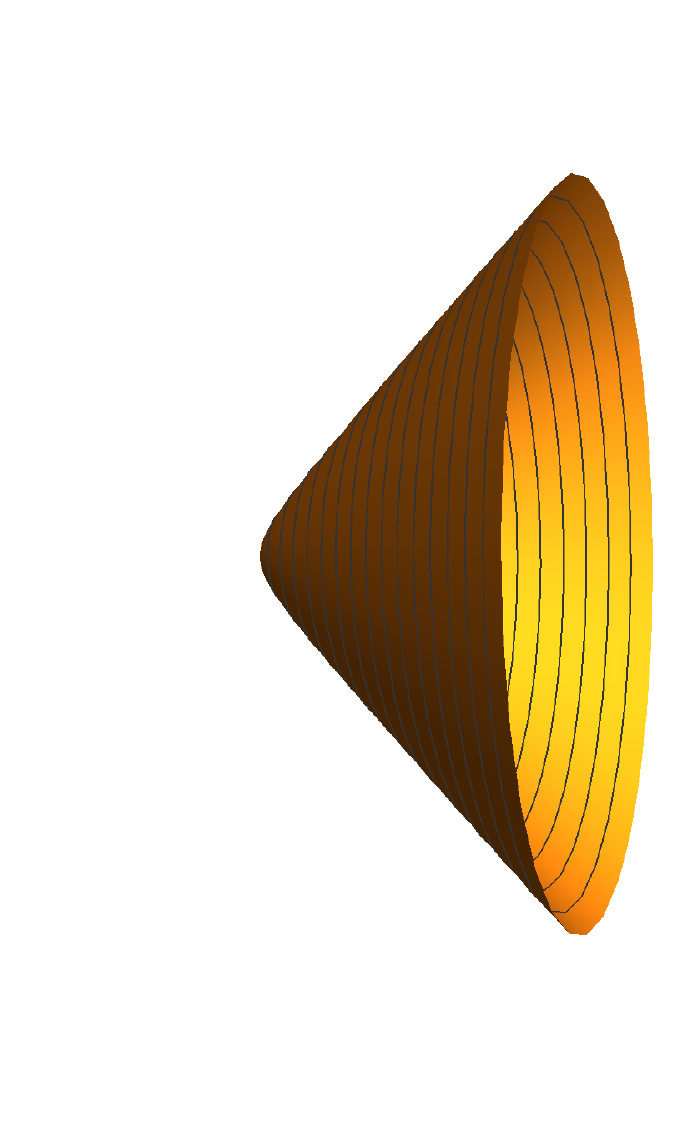}
\caption{$\IC/\IZ_{2}$ with~$r=2$}
\label{figCmodZ2}
\end{figure}

When the FI parameter~$r$ vanishes the metric~\eqref{metricC2Z2r} is exactly that of~$\IC/\IZ_{2}$
(i.e.~as induced from the ambient complex plane). We can read off from the metric~\eqref{metricC2Z2r} that 
the FI parameter smoothes out the singularity near the tip~$M=0$, but its effect dies out near the asymptotic region.
It was observed in \cite{Aharony:2016jki} that the elliptic genus of this model is exactly that of~$\IC$,
prompting the conjecture that this GLSM flows from~$\IC/\IZ_{2}$ (deformed by~$r$ as above) to~$\IC$. 
We shall review the computation of the elliptic genus in the next section.

\vspace{0.2cm}

\ndt {\bf Example 3:}  {\bf $A_{1}$ space}, $(n,d)=(3,2)$. 

Our third example is a four-dimensional manifold modelled by three chiral superfields~$\Phi_{i}$ 
and one gauge superfield. The charges of~$\Phi_{i}$ are~$(1,-2,1)$, respectively. 
The fields~$X=\phi_{1}^{2} \phi_{2}$, $Y=\phi_{2} \phi_{3}^{2}$, $Z=\phi_{1} \phi_{2} \phi_{3}$ 
obey~$XY=Z^{2}$, $X,Y,Z \in \IC$, which is the algebraic equation of the~$A_{1}$ space. 
We can solve the~$D$-term equation 
\be
|\phi_1|^2 - 2|\phi_2|^2+ |\phi_3|^2 -r \= 0
\ee
by writing
\be
\phi_{1}\=\rho \cos \frac{\eta}{2} \, e^{i\theta_{1}} \, , \qquad 
\phi_{2}\=\sqrt{\frac{\rho^{2}-r}{2}}\,  e^{i\theta_{2}}\, , \qquad 
\phi_{3}\=\rho \sin \frac{\eta}{2}  \, e^{i\theta_{3}} \, ,
\ee
where the angle~$\eta$ has a periodicity of~$\pi$. The angles~$\psi_{1}=2\theta_{1}+\theta_{2}$,
$\psi_{3}=2\theta_{3}+\theta_{2}$ are the arguments of the gauge invariant coordinates~$X,Y \in \IC$. 
Implementing the further~$U(1)$ gauging by any of the two methods shown above,
we obtain the tree-level metric.
which coincides with the asymptotic form of the Eguchi-Hanson metric for~$A_{1} = \IC^{2}/\IZ_{2}$. 
The full metric is not that of~$A_{1}$ even when~$r=0$, and only agrees with the ALE metric asymptotically. 
This can be immediately seen, for example, from the fact that the Ricci scalar curvature of this model only 
vanishes asymptotically. 

The elliptic genus of this model was computed in \cite{Harvey:2014nha} and it was found that it agrees with 
that of~$\IC^{2}/\IZ_{2}$ (as we shall  review in the next section). 
This leads to the natural conjecture that this GLSM flows from the metric~\eqref{metricC2Z2r} to~$\IC^{2}/\IZ_{2}$,  
or more precisely the~$A_{1}$ space resolved by the FI parameter~$r$, i.e.~the Eguchi-Hanson metric.  
We comment that~$\CN=(4,4)$ supersymmetric GLSMs for~ALE spaces directly give the ALE 
hyperK\"ahler metric in the UV, and there is no RG flow. This is a reflection of the enhanced 
supersymmetry. 

This example can be easily generalized to all the~$A_{k}$ models with~$k$ gauge multiplets and~$k+2$ 
chiral multiplets with charges~$(1,-2,1,0,\cdots 0)$, $(0,1,-2,1,0, \cdots, 0)$, $\cdots$, $(0, \cdots, 0,1,-2,1)$.
We shall not discuss more details of these models in this paper.

\vspace{0.2cm}

\ndt {\bf Example 4:} {\bf Conifold}, $(n,d)=(4,3)$. 

Our final example is a six-dimensional manifold modelled by four chiral superfields~$\Phi_{i}$ 
and one gauge superfield. The charges of~$\Phi_{i}$ are~$(+1,+1,-1,-1)$, respectively. 
The fields~$X=\phi_{1} \phi_{3} $, $Y=\phi_{2} \phi_{4}$, 
$U=\phi_{1} \phi_{4} $, and $V=\phi_{2} \phi_{3}$ obey the algebraic equation of the conifold~$XY=UV$.
At the level of the metric the pattern is similar to the above two examples. The UV metric is  asymptotically that of the 
conifold, deformed by the FI parameter~$r$ which smoothes out the singularity near the tip. 
The Ricci tensor and Ricci scalar does not vanish when~$r=0$, but approach zero asymptotically. 
We conjecture that this model flows to the conifold.  We present the elliptic genus of this model 
in Section~\ref{sec:EllGenTor}.

\subsection{The squashing deformation \label{sec:Squashed}}

The GLSM~\eqref{action0} has, in addition to the~$U(1)^{n-d}$ gauge symmetry, an independent 
global $U(1)^{d}$ flavor symmetry under which the chiral superfields carry some charges~$F_{i}^{\ell}$, 
$\ell = 1, \cdots, d$. The squashing deformation \cite{Hori:2001ax} involves gauging this symmetry and simultaneously 
introducing a set of~$d$ compensator chiral superfields which translate under the flavor gauge fields. 
The action of the squashed model has three changes compared to the action~$S_{0}$ in~\eqref{action0}. Firstly, one introduces 
a set of flavor gauge fields~$V'_{\ell}$, $\ell = 1, \cdots, d$ with canonical kinetic terms
\be
S_{1} \= - \frac{1}{4\pi e'^{2}} \int\dd^2x\,  \int \dd^{4} \theta \sum_{\ell=1}^{d} \, \bSigma'_{\ell} \, \Sigma'_{\ell} \, . 
\ee
Secondly, the kinetic term of the chiral superfields in~$S_{0}$ undergo the modification:
\be
\sum_{a=1}^{n-d}Q_{i}^{a} V_{a} \; \rightarrow \; \sum_{a=1}^{n-d}Q_{i}^{a} V_{a} \+ 
\sum_{\ell=1}^{n}F_{i}^{\ell} V'_{\ell} \, ,
\ee
to give an action which we call~$S'_{0}$. 
Thirdly, one has the additional term in the action:
\be
S_{2} \= {1\over 2\pi}\int\dd^2x\, \int \dd^4\theta \, \sum_{\ell=1}^{d} \, {k_{\ell}\over 4}(P_{\ell}+\bP_{\ell}+V'_{\ell})^2 \, . 
\ee
The action of the squashed toric model is given by:
\be \label{Ssquashed}
S_\text{squashed} \= S'_{0} + S_{1} + S_{2} \, . 
\ee
The vacuum manifold of this theory is found by solving the constraint equations imposed by setting both $D$ and $D'$ to zero:
\bea
&&D_{a} \= - e_{a}^{2} \, \mu_{a} \, , \qquad  \mu_{a} \= \sum_{i=1}^{n} Q_{i}^{a} \, |\phi_{i}|^{2} - 
r_{a} \, , \qquad a = 1, \dots, n-d \, , \label{Dterm1} \\
&&D'_\ell\=-e'^{2}\mu'_\ell,\qquad \mu'_\ell\=\sum_{i=1}^nF^\ell_i \, |\phi_i|^2+k \, \Re P_{\ell},\qquad \ell=1,...,d\,.
\label{DtermP}
\eea
The physically inequivalent vacua are given by
\be
\wt V\=\mu^{-1}(0)/(U(1)^{n-d}\times U(1)^d)\,, \qquad \mu\=(\mu_a,\mu'_\ell) \, . 
\ee

Thus we see that the vacuum manifolds of the squashed models---the squashed toric manifolds---are also toric manifolds 
with a local~$U(1)^{d}$ action~\cite{Hori:2001ax}. 
We can choose to parameterize the base of the vacuum manifold of the squashed models 
by~$\Re P_{\ell}$ and, by a gauge choice, the circle fibres by~$\Im P_{\ell}$. In the interior of the 
base, where all the original gauge invariant coordinates are non-zero and finite, this circle has a fixed 
radius of order~$\sqrt{k_{\ell}}$, whereas at the edges and corners (where one or more of the original 
gauge invariant coordinates are zero or infinity) some of the circles shrinks to zero size. 
An important difference with the unsquashed case is that even when the sum of the original gauge charges 
are zero, i.e.~when the unsquashed vacuum manifold flows to a Calabi-Yau manifold, the squashed 
deformations break the Ricci-flat condition. Nevertheless the conjecture is that they flow to a~$\CN=(2,2)$ 
SCFT with a non-trivial dilaton profile.

We now illustrate some of the features of the squashed toric manifolds. 
We begin with the K\"ahler form of the free theory which can be read off from the action~\eqref{Ssquashed} to be 
(with~$\phi_{i} = \rho_{i} e^{i\theta_{i}}$ as before):
\be
\wt \o \= \o +\frac{k}{2} \sum_{\ell=1}^{d} k_{\ell} \, \Re p_\ell \wedge \text{Im}\, p_{\ell} \, , \qquad  
\o \=  \sum_{i=1}^{n} \,  \rho_{i} \, d\rho_{i} \wedge d\theta_{i}  \, .
\ee
Here~$\o$ is the K\"ahler form of the unsquashed model written in terms of the original chiral fields before 
any gauging. The quotienting procedure can be done by using the differential of the~$D$-term constraints~\eqref{Dterm1}, \eqref{DtermP} 
to obtain 
\be
\wt \o \=  \sum_{i=1}^{n} \,  \rho_{i} \, d\rho_{i} \wedge d \wt \theta_{i} \, , \qquad 
\wt \theta_{i} \= \theta_{i}- \sum_{\ell=1}^{d} F^{\ell}_{i} \, \text{Im}\, p_{\ell} \, ,
\ee
and then expressing the~$\rho_{i}$ in terms of the gauge-invariant coordinates~$\wt Z_{\ell}$ of the squashed model. 

It may be perhaps more instructive to perform the quotienting in two steps: 
the unsquashed model already comes with a set of 
coordinates~$Z_{\ell} = Z_{\ell} (\{ \phi_{i} \})$, $\ell=1,\cdots,d$ that are gauge-invariant with 
respect to the gauge transformations generated by~$V_{a}$, $a=1,\cdots,n-d$ in terms of which the
K\"ahler form can be written as:
\be
\o \= \sum_{i,\overline{j}=1}^{d} \o_{i \overline{j}} \, dZ_{i} \wedge d \overline{Z}_{\overline{j}} \, . 
\ee
We can now do the squashing deformation, i.e.~the gauging with respect to the fields~$V'_{\ell}$, for which 
a set of fully gauge-invariant coordinates is:
\be
\wt Z_{\ell} \= Z_{\ell} \Bigl(\bigl\{ \phi_{i} \, \exp\bigl(-\sum_{\ell=1}^{d} F_{i}^{\ell} P_{\ell}\bigr) \bigr\}\Bigr) \, , 
\qquad \ell=1,\cdots,d \, ,
\ee
where~$Z_{\ell} (\{ \phi_{i} \})$ are the gauge-invariant composite fields of the unsquashed model.
We write~$Z_{\ell} = R_{\ell} \, e^{i \psi_{\ell}}$ and~$\wt Z_{\ell} = \wt R_{\ell} \, e^{i \wt \psi_{\ell}}$.

We now illustrate this in the simple example of the squashed~$\IC\IP^{1}$ model for which the 
vacuum manifold is:
\be
\quad \wt V(r) \= \{|\phi_1|^2+|\phi_2|^2=r\,,\; F_1|\phi_1|^2+F_2|\phi_2|^2=-k \, \Re P\}/(U(1)\times U(1))\,.
\ee 
The complex coordinate~$Z=(\phi_{1}/\phi_{2}) = R \, e^{i \psi}$ is the gauge-invariant coordinate of the unsquashed
model, and $\wt Z = Z e^{(F_2-F_1)P}$ is the corresponding coordinate of the squashed
model  invariant under both the~$U(1)$ gauge transformations.  
As explained in the previous section, the K\"ahler form of the unsquashed model is 
(using the differential of the D-term constraint):
\be
\o \= \sum_{i=1}^{2} \rho_{i} \, d\rho_{i}  \wedge d \theta_{i} \= \rho_{1} \, d\rho_{1}  \wedge d \psi \, 
\= r \, \frac{R \, dR \wedge d\psi}{(1+R^{2})^{2}} \, .
\ee
The K\"ahler form of the corresponding squashed model is
\be
\wt \o \ = \o +\frac{k}{2} \, k \, \Re p \wedge \text{Im}\, p \= \rho_{1} \, d\rho_{1}  \wedge d \wt \psi 
\= r \, \frac{R \, dR \wedge d\wt \psi}{(1+R^{2})^{2}} 
\ee
where we have used the differential of the $D'$-term constraint in writing the second equality.

We now want to convert everything to~$\wt Z$,~$\overline{\wt Z}$ or, equivalently,~$\wt R$ and~$\wt \psi$,
for which we use the trick of Appendix~\ref{sec:2dtormetric} to convert the coordinate~$R$ to the 
coordiante~$\wt R(R)$. In this manner we find the corresponding metric to be:
\be
ds^{2} \= r \, \frac{R}{(1+R^{2})^{2}} \, \biggl(\frac{\wt R'(R)}{\wt R(R)} dR^{2} +  \frac{\wt R(R)}{\wt R'(R)} 
d\wt \psi^{2} \biggr)\,.
\ee
Computing the derivative we obtain
\be
\frac{1}{\wt R}\frac{d \wt R}{dR} \= \frac{2 (F1 - F2)^2 r R^2 + k (1 + R^2)^2}{k R (1 + R^2)^2}
\= \frac{2 r R^2 + \widehat k (1 + R^2)^2}{\widehat k R (1 + R^2)^2}\, , 
\ee
where $\widehat k=k/(F_1-F_2)^2$. The metric is then given by
\be \label{squashedCP1}
ds^2\= \frac{r(2 r R^{2} + \widehat k (1+R^{2})^{2})}{\widehat k(1+R^{2})^{4}} \, dR^{2} 
+ \frac{\widehat k rR^2}{2 R^2 r \+ \widehat k (1 + R^2)^2} d\psi^{2} \, , 
\ee
which has the shape of a sausage~\cite{Fateev:1992tk, Fendley:1992dm}.

Now we briefly discuss the squashed~$\mathbb C/\mathbb Z_2$  model. 
The vacuum equations for this model are:
\be
\quad \wt V(r)\=\{|\phi_1|^2-|\phi_2|^2=r \, ,\; F_1|\phi_1|^2+F_2|\phi_2|^2=-k\,\text{Re}P\}/(U(1)\times U(1))\,.
\ee 
This is a one dimensional complex space with natural gauge invariant complex coordinate $M=\phi_{1}\phi_{2}\,e^{{-(F_{1}+F_{2})P}}$. 
At a generic point on the Higgs branch labelled by the vacuum expectation value of $M$, both the gauge fields $V_{\mu}$ and $V'_{\mu}$ 
are massive\footnote{The masses of the gauge fields $V_{\mu}$ and $V'_{\mu}$ are of the order of $e$ and $e'\sqrt{k}$, respectively.} 
and therefore, at the energy scale below the masses of the gauge fields, we can integrate them out to get the non linear sigma model. 
Solving for the equations of motion for the gauge fields $V_{\mu}$ and $V'_{\mu}$, one obtains 
\be
V_{\m}\=\frac{r_{1}^{2}\,\p_{\m}\theta_{1}-r_{2}^{2}\p_{\m}\th_{2}+(F_{2}r_{2}^{2}-F_{1}r_{1}^{2})V'_{\m}}{r_{1}^{2}+r_{2}^{2}}\,,
\ee
where
\be
V'_{\mu}\=\frac{(r_{1}^2 + r_{2}^2)\, k\, \p_{\mu}\text{Im}P + 
 2 b_{1}\, r_{1}^2 \,r_{2}^2\, (\p_{\mu}\theta_{1}+ \p_{\mu}\theta_{2})}{2r_{1}^{2}\,r_{2}^{2}\,b_{1}^{2} + (r_{1}^2 + r_{2}^2)\, k}\,.
\ee
Here $\phi_{1}\=r_{1}e^{i\theta_{1}},\,\phi_{2}\=r_{2}e^{i\theta_{2}}$ and $b_{1}\=F_{1}+F_{2}$. 

Substituting the above expressions for the gauge fields 
in the kinetic part of the Lagrangian
\be
 |D_{\mu}\phi_{1}|^{2}+|D_{\mu}\phi_{2}|^{2}+\frac{k}{2}|D_{\mu}P|^{2} \, ,
\ee
and also using the $D$ and $D'$ constraints equations to eliminate~$\text{Re}P$ and~$r_{2}$ in terms of~$r_{1}$ and~$r$, we obtain the tree-level UV metric of the squashed~$\mathbb C/\mathbb Z_2$:
\be \label{squashedC2Z2}
ds^2\=\frac{(2 \rho^2 + \wt k\sqrt{4 \rho^2 + r^2})}{\wt k\,(4 \rho^2 + r^2)} \,   d\rho^2
\+ \frac{\wt k \, \rho^2}{ (2 \rho^2 + \wt k\sqrt{4 \rho^2 + r^2})} \,  d\psi^2\,,
\ee
where $\rho=r_{1}r_{2}\=r_{1}\sqrt{r_{1}^{2}-r}$, and $\psi=\theta_{1}+\theta_{2}-b_{1}\text{Im}P$. The metric depends on the FI parameter, $r$, and~$\wt k$ which is the ratio of~$k$ and~$b^{2}_{1}$,  $\wt k=k/b_1^2$.

\begin{figure}[h]
\label{figSquashedCmodZ2}
\centering
\includegraphics[width=8cm]{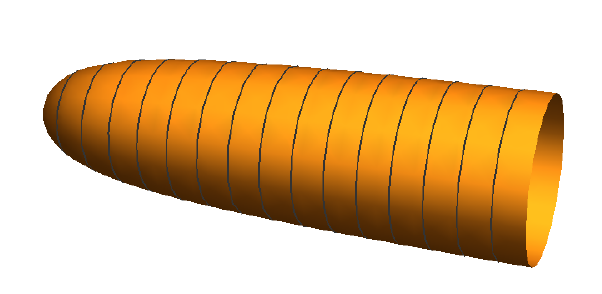}
\caption{Squashed $\IC/\IZ_{2}$ with $\tilde k=1$ and $r=1$}
\end{figure}

The unsquashed vacuum manifold is~$\IC/\IZ_{2}$ which has been smoothed near the tip. 
The squashed manifold, in contrast, has a cigar-like shape. The circle which grows without bound has been
squashed so as to give an asymptotic cylinder.
We shall see in Section \ref{sec:EllGen} that the elliptic genus of the squashed~$\IC/\IZ_{2}$ is exactly 
that of the cigar manifold, based on which we conjecture that the GLSM for squashed $\IC/\IZ_{2}$ 
describes an RG flow from~\eqref{squashedC2Z2} to the cigar 
SCFT\footnote{We note that that the metric~\eqref{squashedC2Z2} is similar to, but slightly different from, the vacuum manifold 
of the model consisting of one chiral field, one gauge field, and one compensator field 
presented in Section 2 of \cite{Hori:2001ax}, which also flows down to the cigar.}.

Similarly the squashed ALE space and the squashed conifold have a shape which can be described as 
a higher-dimensional cigar, i.e.~a~$(2d-1)$-dimensional sphere fibered over the radial direction in a manner 
that the radius of the sphere asymptotes to a constant. These are similar to the manifolds described in \cite{Hori:2002cd}, 
but again the details are slightly different. We conjecture that the IR fixed points where they flow to are 
the same as that of \cite{Hori:2002cd}, namely the IR SCFT reached by the theory on NS5-branes wrapped on certain 
compact manifolds. Our elliptic genus computations of these models in Section~\ref{sec:EllGen}
may thus have possible applications to the gauge theories living on such brane configurations.

\section{Elliptic genus of toric sigma models \label{sec:EllGenTor}}

In this section we review the path integral derivation of the elliptic genus of toric sigma models, 
and illustrate it with the examples discussed in the previous section. 
As  mentioned in the introduction the elliptic genus is the partition function of the GLSM on a 
two-dimensional flat torus with periodic boundary conditions on fermions and bosons, with 
background R-symmetry gauge field~$A^R_\mu$ and background flavor symmetry gauge 
field~$V'_\mu$ which couple to the dynamical fields through covariant derivatives. 

In the previous section we saw that the~$d$-dimensional toric manifold is labelled by the gauge invariant 
complex coordinates~$Z_{\ell}$, $\ell = 1, \cdots, d$. The toric manifold has a non-trivial~$U(1)^{d}$ action 
which is diagonalized by~$Z_{\ell}$ transforming as~$Z_{\ell}\rightarrow e^{i\phi_{\ell}}Z_{\ell}$. 
For each of these~$U(1)$s we will associate a chemical potential $v_{\ell}$ which couples to the 
corresponding conserved current. As we shall see below the elliptic genus of the toric manifold depends 
only on the parameters $\tau$, $z$, and $v_{\ell}$.

We focus on the theories of the type discussed in the previous section, namely $\mathcal N=(2,2)$ 
supersymmetric gauge theory with gauge group $U(1)^{n-d}$ coupled to~$n$ chiral multiplets. 
In order to compute the elliptic genus we need conserved left and right-moving~$U(1)$ R-symmetries. 
These are the theories that obey the Calabi-Yau condition and flow to~$\CN=(2,2)$ SCFTs. From our 
discussion in the previous section, this means that we have to restrict our attention to the non-compact 
models. We shall return to a compact example in Section~\ref{sec:Compact} and compute its Witten index. 

We denote the gauge charges of the chiral multiplets by~$Q_i^{a}$, $i=1,\cdots,n$, $a=1,\cdots, n-d$, 
and the flavor charges by~$F_i^{\ell}$, $\ell = 1, \cdots, d$. 
The computation of the partition function depends on the holonomies
\be
u^{a}\=\oint_A V^{a}-\tau\oint_B V^{a} \, , \qquad u'^{\ell} \=\oint_A V'^{\ell}-\tau\oint_B V'^{\ell} \, ,\qquad 
z\=\oint_A A^R-\tau\oint_B A^R \, ,
\ee
along the two cycles of the torus, which we collectively denote as~$u,u',z,$ respectively. 
The large gauge transformation symmetries of the gauge fields imply\footnote{To be more precise 
the redundancy under large gauge transformations restricts the holonomy of the dynamical gauge field, 
in this case~$u$, to take values in~$\mathbb E_\tau=\mathbb C/(\mathbb Z\tau+\mathbb Z)$, while 
covariance under large gauge transformation of the background gauge fields allows us to 
restrict~$u'$ and~$z$ also to take value in the same torus. In fact in the next section,~$V'$ will become 
dynamical and correspondingly the partition function becomes \emph{invariant} under large 
gauge transformations. } that the complex parameters~$u^{a},u'^{\ell}$, and~$z$ take values 
in~$\mathbb E_\tau=\mathbb C/(\mathbb Z\tau+\mathbb Z)$. The chemical potentials~$v_{\ell}$ 
are linear combinations of the~$u'_{\ell}$s determined by the corresponding gauge-invariant composite 
fields built out of the chiral multiplets~$\phi_{i}$.

The elliptic genus of such a GLSM was computed in \cite{Benini:2013nda,Benini:2013xpa} using 
the technique of supersymetric localization. The localization technique reduces the path integral to 
an integral over the localization manifold which is the set of solutions to the off-shell BPS equations 
of the right-moving supercharge~$Q$. The localization manifold in this case is labelled by arbitrary 
values of the holonomies~$u^{a} \in E_{\t}$ of the gauge fields and all other fields set to zero. The 
classical action of the theory on this manifold vanishes and thus the integrand reduces to a one-loop 
determinant of the quadratic fluctuations of a certain deformation called the localizing 
action. In this case the one loop determinant (with zero modes removed) is
\be
Z_{\text{1-loop}}(\tau,z,u,u')\=\biggl(\frac{-i \, \eta(\tau)}{\vth_1(\tau,z)}\biggr)^{n-d} \, 
\prod_{i=1}^n\frac{\vth_1(\tau,(\frac{R_i}{2}-1)z+Q_i\cdot u
+F_i\cdot u')}{\vth_1(\tau,\frac{R_i}{2}z+Q_i\cdot u+F_i\cdot u')} \, . 
\ee
The first factor in the above expression comes from the one loop computation 
of the vector multiplets and the second factor comes from the chiral multiplets.
Here we have introduced the notation
\be
Q_i\cdot u\=\sum_{a=1}^{n-d} Q^a_i \, u^a \, ,  \qquad F_i\cdot u'\=\sum_{\ell=1}^{d} F^\ell_i \, u'^\ell \, .
\ee
In the above formula~$R_{i}$ is the vector R-charge of the boson in the~$i$th chiral multiplet. 
As we do not have any superpotential in our theories, we can shift~$R_{i}$ by the linear combinations of 
the gauge and flavor symmetries to set it to zero. In the following discussion, we will assume 
that the R-charges of all the chiral multiplets are zero, they can be reinstated easily in all our formulas.

The one loop determinant $Z_{\text{1-loop}}(\tau,z,u,u')$ has poles in $u\in \mathbb C^{n-d}$, along 
certain hyperplanes defined by the condition that one or more chiral multiplets become massless which 
is given by
\be\label{poles1}
Q_{i}\cdot u+F_{i}\cdot u'=0\,\quad \text{mod}\,\,\IZ\tau+\IZ\,.
\ee
The integral over $u$ reduces to computing the residues of $Z_{\text{1-loop}}(\tau,z,u,u')$ at the set of 
poles~$\mathfrak{M}_{\text{sing}}$ where $m$ ($m\geq n-d$) of these hyperplanes intersect. The partition 
function is then given by
\be\label{JKresidue1}
\chi_\text{ell}(M_{\text{tor}};\tau,z,u')\=
-\sum_{u_*\in\mathfrak{M^*}_{\text{sing}}}\underset{u=u_*}{\text{JK-Res}} (Q(u_*),\eta)\,
Z_{\text{1-loop}}(\tau,z,u,u') \, , 
\ee
where $\text{JK-Res}(Q(u_*),\eta)$ is a residue operation in~$(n-d)$ complex dimensions called 
the Jeffrey-Kirwan residue \cite{Jeffrey:9307,Benini:2013xpa}. Here~$\eta$ is an arbitrary vector 
in~$\mathbb R^{n-d}$ needed to define this residue operation, although the final result does not 
depend on the choice of such a vector.  For each~$u \in \mathfrak{M}_{\text{sing}}$, 
$Q(u)$ is the set of (at least~$n-d$) charges defining the hyperplanes intersecting at~$u$. 
The set~$\mathfrak{M^*}_{\text{sing}}$ is the subset of $\mathfrak{M}_{\text{sing}}$ defined by 
the condition that, for any point~$u^{*} \in \mathfrak{M^*}_{\text{sing}}$, the vector~$\eta$ is 
contained in the cone generated by $(n-d)$ linearly independent charge vectors in $Q(u_*)$. 

We will now review the modular and elliptic properties of the expression~\eqref{JKresidue1}. 
We will then illustrate the formula \eqref{JKresidue1} using the examples discussed in 
Section~\ref{sec:GLSM}. The first example $\IC\IP^{1}$ does not have an anomaly-free R-symmetry 
and therefore the expression~\eqref{ellgentorEq} for the elliptic genus is not well defined. 
The other three examples~$\IC/\IZ_{2}$, ALE space, and the conifold do 
have a conserved R-symmetry and we proceed to discuss them in turn.

\subsection{Modular and elliptic properties}
The function $\chi_\text{ell}(M_{\text{tor}};\tau,z,u')$ is holomorphic in $z$ and meromorphic in 
the~$u'^\ell$ variables. We now show that it is a Jacobi form in~$d+1$ elliptic 
variables~$(z,u'^{1}, \cdots, u'^{d})$ (and one modular variable~$\t$) with weight $0$ and index
\be \label{MultiIndex}
M \= 
\begin{pmatrix} 
d/2 & -b_{1}/2 & \cdots & -b_{d}/2\\ 
-b_{1}/2 & 0 & \cdots & 0\\ 
\cdots & 0 & \cdots & 0 \\
-b_{d}/2 & 0 & \cdots & 0\\ 
   \end{pmatrix} \,,
\ee
that is \,\ndt $M_{zz}=d/2$, $M_{zu'_{\ell}}=-b_{\ell}/2$, and the rest of the entries are zero.
For future use we denote the sum of the flavor charges for each flavor symmetry as:
\be
b_{\ell}\=\sum_{i=1}^{n}F^{\ell}_{i} \, .
\ee
For the definition of multivariable Jacobi forms see Appendix \ref{sec:Jacobi}. 

The function $Z_{\text{1-loop}}(\tau,z,u,u') $ transforms as follows:
\begin{enumerate}
\item Under the elliptic transformation of the variable~$z$ with the other elliptic variables~$u'^{\ell}$ fixed 
we have (using the CY condition $\sum_{i=1}^{n}Q_{i}^{a}=0$): 
\be
Z_{\text{1-loop}}(\tau,z+\lambda\tau+\mu,u,u')\=
e^{-2\pi i\left(\frac{d}{2}\lambda^{2}\,\tau+\lambda(z\,d-\sum_{\ell=1}^{d}\,b_{\ell}\,u'^{\ell})\right)}\,
Z_{\text{1-loop}}(\tau,z,u,u'),\quad \l,\mu\in\IZ\,.
\ee
\item Under the elliptic transformation of the variable $u'^{\ell}$ with $z$ fixed\,,
\be
Z_{\text{1-loop}}(\tau,z,u,u'^{\ell}+\lambda^{\ell}\tau+\mu^{\ell})
\=e^{2\pi iz\sum_{\ell=1}^{d}b_{\ell}\,\lambda^{\ell}}\,Z_{\text{1-loop}}(\tau,z,u,u')\,.
\ee
\end{enumerate}

We can now deduce the elliptic transformation of the function~$\chi_\text{ell}(M_{\text{tor}};\tau,z,u')$ 
using these identities. Firstly, under the shift of~$z$ as~$z\rightarrow z+\lambda\tau+\mu$ 
keeping~$u'^{\ell}$ fixed, the locations of the poles in the function~$Z_{\text{1-loop}}(\tau,z,u,u')$ 
remain unchanged. Therefore one can pull out the phase in the JK residue operation to obtain 
\bea\label{elliptictransf1}
&& \chi_\text{ell}(M_{\text{tor}};\tau,z+\lambda\tau+\mu,u')  \nn \\
&& \qquad \qquad \= -\sum_{u_*\in\mathfrak{M^*}_{\text{sing}}}\underset{u=u_*}{\text{JK-Res}} 
(Q(u_*),\eta)e^{-2\pi i\left(\frac{d}{2}\lambda^{2}\tau+\lambda(zd-\sum_{\ell=1}^{d}b_{\ell}u'^{\ell})\right)}
Z_{\text{1-loop}}(\tau,z,u,u')\nn\\
&& \qquad \qquad \= e^{-2\pi i\left(\frac{d}{2}\lambda^{2}\tau+\lambda(zd-\sum_{\ell=1}^{d}
b_{\ell}u'^{\ell})\right)}\chi_\text{ell}(M_{\text{tor}};\tau,z,u')\, .
\eea
Under the other elliptic transformations~$u'^{\ell}$ as~$u'^{\ell}\rightarrow u'^{\ell}+\lambda^{\ell}\tau+\mu^{\ell}$ 
with~$z$ fixed. The locations of poles of~$Z_{1-\text{loop}}$ change from \eqref{poles1} to
\be
Q_{i}(u_{*})+F_{i}(u')+\sum_{\ell=1}^{d}F^{\ell}_{i}(\lambda^{\ell}\tau+\mu^{\ell})\=0\quad \text{mod}\,\,\IZ\tau+\IZ\,.
\ee
Since~$F_{i}^{\ell}\in\IZ$, we see that the set of poles in~$E_{\t}=\IC/\IZ\tau+\IZ$ does not change, 
although the individual poles may get rearranged. Thus we get
\bea\label{elliptictransf2}
\chi_\text{ell}(M_{\text{tor}};\tau,z,u'^{\ell}+\lambda^{\ell}\tau+\mu^{\ell})\=e^{2\pi iz\sum_{\ell=1}^{d}b_{\ell}\lambda^{\ell}}\chi_\text{ell}(M_{\text{tor}};\tau,z,u')\,.
\eea
As we will see later this condition is essential in order to gauge the flavor symmetry.

The modular transformation property of~$Z_\text{1-loop}$ applied to Formula~\eqref{JKresidue1} yields the 
modular transformation of~$\chi_\text{ell}(M_{\text{tor}})$. Putting this together with the above elliptic transformation 
properties, we see that~$\chi_\text{ell}(M_{\text{tor}};\tau,z,u')$ is a Jacobi form with weight~$0$ 
and index~$M$.

\subsection{Examples \label{sec:ellgenexamples}}

We now illustrate the considerations above in the examples that we introduced in Section~\ref{sec:GLSM}. 
As explained above the elliptic genus is only well-defined for the non-compact examples. 
In each case the elliptic genus is a Jacobi form holomorphic in~$z$ and meromorphic in~$v_{\ell}$, $\ell=1,\cdots,d$.
This is consistent with the fact that~$v_{\ell}$ is chemical potential for the rotation of the 
gauge-invariant complex variable~$Z_{\ell}$. The basic model for the meromorphicity in the flavored elliptic 
genus is the complex plane~$\IC$. 
The pole arises from the bosonic zero mode on the plane. The divergence is 
regularized by the chemical potential for flavor rotation, i.e.~the angular momentum of the plane
whose fixed point is the origin of~$\IC$. The poles in our expressions can thus be associated with  
the fixed points of the~$U(1)^{d}$ symmetries. 

We note that the FI term which smooths out the orbifold singularities (see e.g.~the discussion around  
Figure~\ref{figCmodZ2} for the case~$\IC/\IZ_{2}$) is a~$Q$-exact term and therefore 
does not affect our computation of the elliptic genus.

\subsection*{$\mathbb C/\IZ_2$ (flowing to $\IC$)}

We consider the $U(1)$ gauge theory with 2 chiral multiplets~$\Phi_1$ and~$\Phi_2$ with gauge 
charges $1,-1$, respectively and flavor charges $F_{1}, F_{2}$, respectively. This model describes 
the space $\mathbb C/\IZ_2$ which is expected to flow to~$\IC$. 
In the $U(1)$ case the Jeffrey-Kirwan residue operation in~\eqref{JKresidue1} reduces to collecting 
the residue from the poles~$u_{i}$ satisfying the condition $\eta Q>0$, where $Q$ is the charge of 
the chiral multiplet becoming massless at $u$. Therefore, choosing the vector $\eta>0$, we obtain
\be
\chi_\text{ell}(\mathbb C;\tau,z,u')\=-\sum_{u_j\in \mathfrak{M}_{\text{sing}}^+}\oint_{u=u_j} \dd u
\frac{i\eta(q)^3}{\vth_1(\tau,-z)}\frac{\vth_1(\tau,-z+u+F_{1}u')}{\vth_1(\tau,u+F_{1}u')}
\frac{\vth_1(\tau,-z-u+F_{2}u')}{\vth_1(\tau,-u+F_{2}u')}\,.
\ee
Here~$F_{i}$ are flavor charges of the chiral multiplets. In the present case, we have only one pole, and for a 
generic value of $u'$ the location of the pole is given by $u+F_{1}u'=0 \, (\text{mod}\,\,\mathbb Z+\tau\mathbb Z)$. 
Using the residue formula \eqref{etatheta} we obtain
\be \label{EllGenC}
\chi_\text{ell}(\mathbb C;\tau,z,u')\=\frac{\vth_1(\tau,-z+(F_1+F_2)u')}{\vth_1(\tau,(F_1+F_2)u')}
\=\frac{\vth_1(\tau,-z+v)}{\vth_1(\tau,v)}\,,
\ee
where~$v=b_{1}u'$, with $b_{1}=(F_1+F_2)$, the flavor charge of the gauge invariant 
variable~$Z=\phi_{1}\phi_{2}$. The expression~\eqref{EllGenC} is equal to the elliptic genus of the~$c=3$ 
superconformal field theory on~$\IC$\,, with a chemical potential for the rotation of the complex 
coordinate~$Z \in \IC$. The function $\chi_\text{ell}(\mathbb C;\tau,z,u')$ is a meromorphic Jacobi form,  
with only one pole in~$E_{\t}$ at~$v=0$, of weight $0$ and index 
\be
M_{\IC} \= \begin{pmatrix} 1/2 & -b_{1}/2 \\ -b_{1}/2 & 0 \end{pmatrix} \, .
\ee

\subsection*{$\text{Resolved}\,A_{1}\text{-Space}$}
We consider the~$U(1)$ gauge theory with 3 chiral multiplets~$\Phi_1$, $\Phi_2$, and~$\Phi_3$ with 
charges $1,-2,1$, respectively. There is a~$U(1)^{2}$ flavor symmetry under which the chiral multiplets~$\Phi_{i}$ 
have charges~$F_{i}^{\ell},\, (\ell=1,2; \,i=1,2,3)$.  
This model describes the $ALE$ space of type~$A_1$. 
We compute the Jeffrey-Kirwan residue by choosing the vector $\eta>0$, for which we pick up residues at poles $u=-F_{1}\cdot u'$ and $u=-F_{3}\cdot u'$. The elliptic genus is then given by 
\bea\label{ellipgenusA1}
\chi_\text{ell}(A_{1};\tau,z,u')&\=&\frac{\vth_1(\tau,-z+(F_3-F_1)\cdot u')}{\vth_1(\tau,(F_3-F_1)\cdot u')}\frac{\vth_1(\tau,-z+(2F_1+F_2)\cdot u')}{\vth_1(\tau,(2F_1+F_2)\cdot u')}\\
&&\qquad+\frac{\vth_1(\tau,-z+(F_1-F_3)\cdot u')}{\vth_1(\tau,(F_1-F_3)\cdot u')}\frac{\vth_1(\tau,-z+(2F_3+F_2)\cdot u')}{\vth_1(\tau,(2F_3+F_2)\cdot u')}\nn\\&\=&\frac{\vth_1(\tau,-z+v_{2}-v_{1})}{\vth_1(\tau,v_{2}-v_{1})}\frac{\vth_1(\tau,-z+2v_{1})}{\vth_1(\tau,2v_{1})}+\frac{\vth_1(\tau,-z+v_{1}-v_{2})}{\vth_1(\tau,v_{1}-v_{2})}\frac{\vth_1(\tau,-z+2v_{2})}{\vth_1(\tau,2v_{2})} \,.\nn
\eea
Here $2v_{1}=(2F_{1}+F_{2})\cdot u'$ and $2v_{2}=(2F_{3}+F_{2})\cdot u'$, which are the chemical potentials 
coupling to the rotations of the gauge invariant 
variables~$X=\phi_{1}^{2}\phi_{2}$, and~$Y=\phi_{3}^{2}\phi_{2}$. The Witten index, i.e.~the value of the elliptic genus~\eqref{ellipgenusA1} 
at~$z=0$ equals 2, which is exactly the Euler character of the Eguchi-Hanson space \cite{Eguchi:1978gw}.
The function~$\chi_\text{ell}(A_{1};\tau,z,u')$ is a meromorphic Jacobi form of weight $0$ and index 
\be
M_{A_{1}} \= \begin{pmatrix} 1 & -b_{1}/2 &-b_{2}/2 \\ -b_{1}/2 & 0 &0\\-b_{2}/2&0&0\end{pmatrix} 
\ee
in the variables $(z,u_{1}',u_{2}')$ with poles at~$v_{1}(\{u'_{\ell}\})=0$ and $v_{2}(\{u'_{\ell}\})=0$.

\subsection*{Conifold}
We consider the~$U(1)$ gauge theory with 4 chiral multiplets~$\Phi_1$,$\Phi_2$,$\Phi_3$ and~$\Phi_4$ 
with charges $1,1,-1,-1$, respectively. 
This model describes the resolved conifold. There is a $U(1)^{3}$ flavor symmetry under which the 
chiral multiplets~$\Phi_{i}$ have charges~$F_{i}^{\ell},\, (\ell=1,2,3; \,i=1,2,3,4)$. Computing the 
Jeffrey-Kirwan residue as in the previous case and picking up the residues at 
the poles for $\eta<0$ (corresponding to the negative charge), we obtain 
\bea
\chi_\text{ell}(\text{Conifold};\tau,z,u')&\=&-\frac{\vth_1(\tau,-z+v_{1})}{\vth_1(\tau,v_{1})}\frac{\vth_1(\tau,-z+v_{3})}{\vth_1(\tau,v_{3})}\frac{\vth_1(\tau,-z+v_{2}-v_{1})}{\vth_1(\tau,v_{2}-v_{1})}\\&&
\quad-\frac{\vth_1(\tau,-z+v_{2})}{\vth_1(\tau,v_{2})}\frac{\vth_1(\tau,-z+v_{2}-v_{1}+v_{3})}{\vth_1(\tau,v_{2}-v_{1}+v_{3})}\frac{\vth_1(\tau,-z+v_{1}-v_{2})}{\vth_1(\tau,v_{1}-v_{2})}\,.\nn
\eea
Here $v_{1}=(F_{1}+F_{3})\cdot u'$, $v_{2}=(F_{1}+F_{4})\cdot u'$ and $v_{3}\=(F_{2}+F_{3})\cdot u'$, which 
are the chemical potentials coupling to the rotations of the gauge invariant 
variables~$X=\phi_{1}\phi_{3}$, $U=\phi_{1}\phi_{4}$ and~$V=\phi_{2}\phi_{3}$. 
Again we see that the the elliptic genus is a meromorphic Jacobi form with index given by~\eqref{MultiIndex}.
The poles are at~$v_{\ell}=0$, $\ell=1,2,3$, and~$v_{4}=v_{2}+v_{3}-v_{1}=0$. These four poles correspond 
to the fixed points of~$X$, $V$, $U$, and~$Y=\phi_{2}\phi_{4}$ which obey~$XY=UV$.

\section{Elliptic genus of squashed toric sigma models \label{sec:EllGen}}

In this section we compute the elliptic genus of the squashed toric sigma models discussed in Section~\ref{sec:GLSM},
and derive the main formula of the paper.  
The starting point is the unsquashed theory discussed in Section~\ref{sec:EllGenTor}, namely 
the~$U(1)^{n-d}$ gauge theory coupled to $n$ chiral multiplets. This theory has a~$U(1)^{d}$ flavor symmetry, 
under which the chiral multiplets have charges $F_i^\ell$, $(i=1,\cdots,n, \ell=1,\cdots,d)$. 
The squashing corresponds to gauging the~$U(1)^{d}$ symmetry and introducing~$d$ compensator~$P$-fields,
as discussed in Section~\ref{sec:Squashed}. The action of the theory is given in Equation~\eqref{Ssquashed}. 
The details of the Lagrangian in the component form is given in Appendix~\ref{glsmLagr}.

We use localization to compute the supersymmetric partition function of these theories. 
The main idea of the computation is very close to that of~\cite{Benini:2013nda}, \cite{Murthy:2013mya}, 
which we follow. 
The first step is to deform the action by a~$Q$-exact term with a coupling~$\lambda$. The~$Q$-exactness implies 
that the answer is independent of the coupling and we can therefore evaluate the path integral in the 
limit~$\l \to \infty$. 
In this limit the path-integral reduces to the critical points of the deformation action, the localization manifold, and the 
computation of its one-loop determinant at these points. 
We choose the deformation such that the localization manifold is the set of solutions to the off-shell BPS equations of 
the supercharge~$Q$ on the vector and chiral 
multiplets.
This is given by the constant gauge field holonomies along the two cycles of~$T^2$. As in the previous section we 
denote the holonomies of the~$U(1)^{n-d}$ gauge fields and the $U(1)^d$ flavor gauge fields by 
$\{u^a\}_{a=1,..,n-d}$ and $\{u'^\ell\}_{\ell=1,..,d}$, respectively. 
The full functional integral reduces to an integral over these holonomies as well as the full field space of the compensator 
multiplets. Noting that the full Lagrangian for the $P$-multiplet fields evaluated on this above localization locus is 
quadratic, we can simply perform the Gaussian path-integral.

The one loop determinant coming from the integration over non zero modes for each vector multiplet is identical and is 
given by $\dfrac{i\eta(\tau)^3}{\vartheta_1(\tau,-z)}$, thus giving the following total contribution from the non zero modes 
of~$n$ vector multiplets:
\be
 Z_{\text{vec}}(\tau,z)\=\left(-\frac{i\eta(\tau)^3}{\vartheta_1(\tau,z)}\right)^n\,.
\ee
However this is not the complete result for the vector multiplet as there are zero modes of kinetic operator for 
gaugino fields $(\bar\lambda_{a+}^{(0)},\lambda_{a+}^{(0)})$ and $(\bar\lambda^{'(0)}_{\ell+},\lambda_{\ell+}^{'(0)})$. 
The zero modes $(\bar\lambda_{a+}^{(0)},\lambda_{a+}^{(0)})$ couple to chiral multiplets through Yukawa interactions 
and the zero modes $(\bar{\lambda}^{'(0)}_{\ell+},\lambda_{\ell+}^{'(0)})$ couple to both chiral and P-multiplets. 
As we see below 
the zero modes $(\bar{\lambda'}^{(0)}_+,\lambda_+^{'(0)})$ are absorbed by the zero modes of the $P$-multiplet 
fermions $(\bar\chi_{\ell-}^{(0)},\chi_{\ell-}^{(0)})$ and the fermion zero modes $(\bar\lambda_{a+}^{(0)},\lambda_{a+}^{(0)})$ 
are absorbed by the Yukawa interactions of chiral multiplets. 
For details of the Lagrangian involving chiral, vector and $P$-multiplets, see Appendix \ref{glsmLagr}.

We first integrate over the fields in the $P_{\ell}$-multiplets whose Lagrangian is:
\bea
\frac{k_\ell}{2}\Big(-D^{'\mu}\bar p_\ell D'_\mu p_\ell+i\bar\chi_{\ell-}(\p_0+\p_1)
\chi_{\ell-}+i\bar\chi_{\ell+}(\p_0-\p_1)\chi_{\ell+}+iD'_\ell(p_\ell+\bar p_\ell)+|F_{p_\ell}|^2\nn\\
-|\sigma'_\ell|^2+i\chi_{\ell+}\lambda'_{\ell-}-i\chi_{\ell-}\lambda'_{\ell+}
+i\bar\chi_{\ell+}{\bar\lambda'}_{\ell-}-i\bar\chi_{\ell-}{\bar\lambda}'_{\ell+}\Big)\,.
\eea
The  integration over the non zero modes of $\chi_{\ell\pm}$ gives 
\be
\prod_{m,n\in \IZ}(m\tau+n+z)\prod_{m,n\in\IZ\setminus(0,0)}(n+m\bar\tau)\,,
\ee 
and the integration over the non zero modes of $\text{Re}\,p_{\ell}$ gives
\be
\prod_{m,n\in\IZ\setminus(0,0)}|(n+m\tau)|\,.
\ee
Next we need to integrate over $\text{Im} \, p_\ell$. 
Since $\text{Im}\,p_\ell $ lives on a circle of unit radius, its mode expansion is given by
\be
\text{Im}\,p_\ell\=2\pi(w_\ell\sigma_1+m_\ell\sigma_2)+\text{Im}\,p_{\ell\text{oscil}}, 
\ee
where $\sigma_{1,2}$ are coordinates on $T^2$ with range~$0\leq \sigma_{1,2}<1$ and~$(m_\ell,w_\ell)$ 
are arbitrary integers. Integrating over the oscillator modes of $\text{Im}\,p_\ell$ gives 
\be
\prod_{m,n\in\IZ\setminus(0,0)}|(n+m\tau)|\,.
\ee
Thus the complete one loop determinant coming from the non zero modes for each $\ell$ is
\be
\frac{\prod_{m,n\in \IZ}(m\tau+n+z)\prod_{m,n\in\IZ\setminus(0,0)}(n+m\bar\tau)}{\prod_{m,n\in\IZ\setminus(0,0)}|(n+m\tau)|^{2}}\=\frac{1}{2\pi}\frac{\vth_{1}(\t,z)}{\eta(\t)^{3}}\,.
\ee
Next we integrate over the zero modes. Integrating over the zero modes $({\bar\lambda}_{\ell+}^{'(0)},\lambda_{\ell+}^{'(0)})$ and $(\bar\chi_{\ell-}^{0},\chi^{(0)}_{\ell-})$ 
gives the factor~$k_\ell^2/4$. The integration over the zero mode $\text{Re}\,p^{(0)}_{\ell}$ 
gives~$\delta(D'_\ell)/k_\ell\tau_2$ for each $\ell$. Integrating over the zero mode of the oscillator part of $\text{Im}\,p_\ell$ gives $2\pi$. 

Putting all this together, we see that the one loop contribution of the fields in a given $P$-multiplet 
is\footnote{We have fixed the overall normalization  of $Z_{P_\ell}(\tau,z,u')$, by comparing to the result of 
Witten index for squashed $\CP^{1}$\,. }
(with~$b_\ell=\sum_{i=1}^nF_i^\ell$):
\be\label{integPfield.1}
Z_{P_\ell}(\tau,z,u')\=\frac{k_\ell}{\tau_2}
\frac{i\vartheta_1(\tau,z)}{\,\eta(\tau)^3} \, \delta(D'_\ell) \sum_{m_\ell,w_\ell\in\mathbb{Z}}
e^{-\frac{\pi k_\ell}{\tau_2}(w_\ell\tau+m_\ell+u'+\frac{b_\ell z}{k_\ell})(w_\ell\bar\tau+m_\ell+\bar{u'}+
\frac{b_\ell z}{k_\ell})} \, . 
\ee
We note that the $k_\ell$-dependence of the non-zero modes cancel out among themselves, and 
the only $k_\ell$-dependence in the above partition function comes from zero modes.

Now we integrate over the fields in the chiral multiplets. We note that in the presence of the $P$-field, the 
correct gauge field background is $V'^{\ell}_{\mu}+\p_{\mu}\text{Im}P_{\ell}$. The chiral multiplet 
determinant in the background of $d$ number of $P$-fields is given by
\be
Z^{(\{w_l\},\{m_l\})}_{\text{chiral}}(\tau,z,u,u',\widehat D)\=\prod_{i=1}^n\prod_{r_i,n_i\in \IZ}
\frac{\text{Num}(\{r_i,n_i\},u,u',\tau,\bar\tau)}{\text{Denom}(\{r_i,n_i\},u,u',\tau,\bar\tau,\widehat D)}\,,
\ee
where
\bea
\text{Num}(\{r_i,n_i\},u,u',\tau,\bar\tau)\=\Big(r_i+n_i\tau+z-Q_i\cdot u-F_i\cdot(u'+w\tau+m)\Big)\times\nn \\
\Big(r_i+n_i\bar\tau+Q_i\cdot \bar u+F_i\cdot(\bar{u'}+w\bar\tau+m)\Big)\,,
\eea
and
\be
\text{Denom}(\{r_i,n_i\},u,u',\tau,\bar\tau,\widehat D)
\=\Big|r_i+n_i\tau+Q_i\cdot u+F_i\cdot(u'+w\tau+m)\Big|^2+i\widehat D_i\,.
\ee
In the case when both $\{D\}$ and $\{D'\}$ are zero we get
\bea
Z^{(\{w_l\},\{m_l\})}_{\text{chiral}}(\tau,z,u,u')&\=&e^{2\pi iz \sum_{i=1}^n F_i\cdot w}
\prod_{i=1}^n\frac{\vartheta_1(\tau,-z+Q_i \cdot u+F_i\cdot u')}{\vartheta_1(\tau, Q_i \cdot u+ F_i\cdot u')}\\
&\=&e^{2\pi i z\sum_{\ell=1}^d b_{\ell}w_{\ell} }Z_{\text{chiral}}(\tau,z,u,u') \, .
\eea
Now we need to integrate over the zero modes of the gaugini $(\lambda^{(0)}_{a+},\bar\lambda^{(0)}_{a+})$ 
and~$\{D\}$ and~$\{D'\}$. We first integrate over~$D'_\ell$'s. Using (\ref{integPfield.1}), we can set~$\{D'\}$ 
equal to zero in the integrand. After setting $D'_\ell=0$, the rest of the integrals can be performed as 
in~\cite{Benini:2013nda,Benini:2013xpa}. 

The result for the complete one loop determinant is:
\bea \label{Z1loop}
&& Z_{\text{1-loop}}(\tau,z,u,u')\=Z_{\text{vec}}(\tau,z)Z_{\text{chiral}}(\tau,z,u,u')\prod_{\ell=1}^d\frac{k_\ell}{\tau_2}
\frac{i\,\vartheta_1(\tau,z)}{\eta(\tau)^3} \times \\
&&  \qquad  \qquad  \qquad  \qquad  \qquad  \qquad 
\times \sum_{m_\ell,w_\ell\in\mathbb{Z}}e^{2\pi i zb_{\ell}w_{\ell}}
e^{-\frac{\pi k_\ell}{\tau_2}(w_\ell\tau+m_\ell+u'+\frac{b_\ell z}{k_\ell})(w_\ell\bar\tau+m_\ell+\bar{u'}+\frac{b_\ell z}{k_\ell})} \,.\nn
\eea
Thus the full partition function is given by
\be
\chi_\text{ell}(\wt M_\text{tor}; \tau,z)\=-\int_{{E_\tau^{d}}} \prod_{\ell=1}^d
\dd^2u'_\ell\sum_{u\in\mathfrak{M^*}_{\text{sing}}}\underset{u=u_*}{\text{JK-Res}}(Q(u_*),\eta)
\, Z_{\text{1-loop}}(\tau,z,u,u') \, . 
\ee
Putting together Equations \eqref{JKresidue1} and \eqref{Z1loop}, 
this can be written as
\bea\label{SquashElliptic2}
\chi_\text{ell}(\wt M_\text{tor}; \tau,z)  \= 
\bigintsss_{E_\tau^{d}} \, \prod_{\ell=1}^d   \frac{\dd^2 u'_\ell}{\t_{2}} \,  
\wt H_\ell (\tau, z, u'_\ell)  \; \chi_\text{ell}(M_\text{tor};\tau,z, u') \, ,
\eea
where we have defined the function 
\be 
\wt H_\ell (\tau, z, u) \= k_{\ell} \sum_{m,w \, \in \, \mathbb{Z}} \,
e^{2\pi i b_\ell w z - \frac{\pi k_\ell}{\tau_2} \bigl( w\tau+m+u+\frac{b_\ell z}{k_\ell} \bigr)
\bigl(w \bar{\tau}+m+\overline{u}+\frac{b_\ell z}{k_\ell } \bigr)} \,,
\ee 
which depends on the models only through the parameter~$b_{\ell}=\sum_{i=1}^{n} F^{\ell}_{i}$.
As we discuss below, the integrand in Equation \eqref{SquashElliptic2} is invariant under 
elliptic transformations of each~$u'_{\ell}$ and therefore the integral is well-defined.

We note that the integrand in \eqref{SquashElliptic2} has poles at $v_{\ell}=0$ coming from the 
term $\chi_\text{ell}(M_\text{tor};\tau,z, u')$. In order to the define the integral, we cut out a small disk 
around each pole (as in~\cite{Harvey:2013mda}),
and then take a limit where the size of the disk goes to zero. For any simple pole (say around~$u=0$) 
the simple zero of the measure factor~$\text{d}^{2}u$ cancels it and therefore this limit is well-defined. 
One may worry that  one gets higher order poles in~$\chi_\text{ell}(M_\text{tor};\tau,z, u')$ (which would 
make the resultant integral over the torus ill-defined), but this does not happen because there are~$d$ 
linearly-independent circles in the geometry, each of which is associated with a flavor symmetry, and 
the locations of the poles are precisely where these circles shrink to zero size.

Now we  comment on the total number of parameters that enter the elliptic genus of our squashed toric sigma models. 
The parameters that enter the formula~\eqref{SquashElliptic2} are~$k_{\ell}$ and~$b_{\ell}$ (for each function~$\wt H_{\ell}$),
and the parameters entering~$\chi_{\ell}(M_\text{tor})$. 
As we observed earlier, the function~$\chi_\text{ell}(M_\text{tor};\tau,z, u')$ depends on $u'$ only through the 
combinations~$v_{\ell}$, $\ell=1, \cdots, d$.  The $v_{\ell}$'s are the chemical potentials associated with the 
gauge invariant variables~$Z_{\ell}$ which are fixed for a given model. The parameters in the gauged model 
are the charges of each~$Z_{\ell}$ under the~$U(1)^{d}$ flavor gauge field, which are~$d^{2}$ parameters.
So it looks like we have a total of~$d^{2}+2d$ parameters but, as we now see, there are actually fewer parameters.

Firstly there are certain scaling transformations of the above parameters which are symmetries of the equation~\eqref{SquashElliptic2}. 
To see this, we use an equivalent expression for the elliptic genus \eqref{unfoldedEq1} which is given as
an integral over the entire complex plane. This expression is invariant under the following rescalings:
\be\label{scaling}
F_{i}^{\ell}\rightarrow \l_{\ell} \,F^{\ell}_{i} \quad \bigl(\Rightarrow\; b_{\ell}\rightarrow \l_{\ell} \, b_{\ell} \bigr) ,
\qquad k_{\ell}\rightarrow \l_{\ell}^{2}\, k_{\ell}, \qquad \forall \, (i,\ell) \, ,
\ee
as can be seen by changing the integration variable from $u'_{\ell}$ to $u'_{\ell}/\l_{\ell}$. 
This reduces the number of independent parameters by~$d$.
Next, the CY conditions $\sum_{i=1}^{n}Q_{i}^{\ell}=0$ imply that the product of all the chiral multiplets 
is gauge invariant and therefore can be expressed as a monomial in the~$d$ gauge invariant coordinates 
of the target space manifold. This gives another~$d$ relations between~$b_{\ell}$'s and certain combinations 
of flavor charges of the gauge invariant degrees of freedom. 
Thus the total number of independent parameters are~$d^{2}$. 

By choosing~$\l_{\ell}=1/b_{\ell}$ in the scaling~\eqref{scaling}, and by using the fact that the 
replacement~$F_{i}^{\ell}\rightarrow \l_{\ell} \,F^{\ell}_{i}$ is completely equivalent 
to~$u_{i}^{\ell}\rightarrow \l_{\ell} \, u^{\ell}_{i}$
in the function~$\chi_\text{ell}(M_\text{tor};\tau,z, \{u'_{\ell} \})$, 
we can rewrite Equation~\eqref{SquashElliptic2} as 
\bea\label{SquashElliptic3}
\chi_\text{ell}(\wt M_\text{tor}; \tau,z)  \= 
\bigintsss_{E_\tau^{d}} \, \prod_{\ell=1}^d   \frac{\dd^2 u'_\ell}{\t_{2}} \,  
H_{\wt k_\ell} (\tau, z, u'_\ell)  \; \chi_\text{ell}(M_\text{tor};\tau,z, \{u'_{\ell}/b_{\ell}\}) \, ,
\eea
where now the function 
\be 
H_k (\tau, z, u) \= k \sum_{m,w \, \in \, \mathbb{Z}} \,
e^{2\pi i w z - \frac{\pi k_\ell}{\tau_2} \bigl( w\tau+m+u+\frac{z}{k} \bigr)
\bigl(w \bar{\tau}+m+\overline{u}+\frac{z}{k} \bigr)} 
\ee 
is a universal function, independent of any features of the models. 
This is the equation that we have presented in the introduction.

\subsection{Modularity and holomorphic anomaly \label{sec:HolAnom}}
We now discuss the elliptic and modular properties of the function~$\chi_\text{ell}(\wt M_\text{tor}; \tau,z)$\,.
We begin with the elliptic property of~$\wt H_\ell$ under the transformation of~$u'$ as~$u'+\lambda\t+\mu$ for $\l\,,\m\in\IZ$\,. The function~$\wt H_\ell$ transforms as 
\bea
\wt H_\ell (\tau, z, u+\l\t+\m) &\=& k_{\ell} \sum_{m,w \, \in \, \mathbb{Z}} \,e^{2\pi i b_\ell w z - \frac{\pi k_\ell}{\tau_2} \bigl( w\tau+m+u+\l\t+\m+\frac{b_\ell z}{k_\ell} \bigr)\bigl(w \bar{\tau}+m+\overline{u}+\l\bar\t+\m+\frac{b_\ell z}{k_\ell } \bigr)} \,, \nn\\
&\=&e^{-2\pi ib_{\ell}\l z} \, \wt H_\ell (\tau, z, u)\,.
\eea
Combining this with the elliptic properties  \eqref{elliptictransf2} of~$\chi_\text{ell}(M_\text{tor};\tau,z, u')$, we see that the integrand in \eqref{SquashElliptic2} is invariant under elliptic transformations. 

In order to compute the modular properties of the above function, it is useful to unfold the integral over $E_\tau$ for each $\ell$ to the entire complex plane. Using the elliptic properties of $\wt H_\ell$ and $\chi_\text{ell}(M_\text{tor};\tau,z, u')$, the Equation \eqref{SquashElliptic2} can rewritten as follows
\bea\label{unfoldedEq1}
\chi_\text{ell}(\wt M_\text{tor}; \tau,z)  \= (\prod_{\ell=1}^d k_{\ell})
\bigintsss_{\IC^{d}} \, \prod_{\ell=1}^d   \frac{\dd^2 u'^{\ell}}{\t_{2}} \, 
e^{ - \frac{\pi k_\ell}{\tau_2} \bigl( u'^{\ell}+\frac{b_\ell z}{k_\ell} \bigr)
\bigl(\overline{u'^{\ell}}+\frac{b_\ell z}{k_\ell } \bigr)} \, 
\chi_\text{ell}(M_\text{tor};\tau,z, u') \,.
\eea
We now have that, under modular transformations,
\bea
\chi_\ell  \Bigl( \wt M_\text{tor}; -\frac{1}{\tau},\frac{z}{\tau}\Bigr)  &\=& (\prod_{\ell=1}^d k_{\ell})
\bigintsss_{\IC^{d}} \, \prod_{\ell=1}^d   \frac{\dd^2 \wt u^{\ell}}{\t_{2}} \, 
e^{ - \frac{\pi k_\ell}{\tau_2} \bigl( \frac{\wt u^{\ell}}{\t}+\frac{b_\ell z}{\t k_\ell} \bigr)
\bigl(\frac{\overline{\wt u^{\ell}}}{\bar\t}+\frac{b_\ell z}{\t k_\ell } \bigr)} \, 
\chi_\text{ell} \Bigl(M_\text{tor};-\frac{1}{\tau},\frac{z}{\t}, \frac{\wt u}{\t} \Bigr) \, ,\nn\\
&\=& e^{\frac{2\pi i}{\t}z^{2}(\frac{d}{2}+\sum_{\ell=1}^{d}\frac{1}{\wt k_{\ell}})}\chi_\text{ell}(\wt M_\text{tor}; \tau,z) \,.
\eea
We use the change of variables~$u'^{\ell}=\wt u^{\ell}/\t$ and the fact that $\frac{d^2u^{\ell}}{\t_{2}}$ is invariant 
under $\t\rightarrow-1/\t$ and $u^{\ell}\rightarrow u^{\ell}/\t$ to obtain the first equality. 
To obtain the second, we use the modular properties of~$\chi_\text{ell}(M_\text{tor};\tau,z, u')$.
Thus we find that~$\chi_\text{ell}(\wt M_\text{tor}; \tau,z)$ is a (one-variable) Jacobi form of weight zero and index 
(recalling that~$\wt k_{\ell} = k_{\ell}/b_{\ell}^{2}$):
\be
m \= \frac{d}{2}+\sum_{\ell=1}^{d} \frac{1}{\wt k_{\ell}}\, .
\ee
This formula for the index is consistent with the conjecture that the GLSM flows to a SCFT with~$c=6m$.

Now we will derive the holomorphic anomaly equation for the elliptic genus of squashed toric 
manifolds~$\chi_\text{ell}(\wt M_\text{tor}; \tau,z)$ following the treatment 
in~\cite{Harvey:2013mda,Murthy:2013mya}. 
We begin by noting that the function~$\wt H_\ell$ satisfies the following heat equation:
\be\label{anomalyEqu}
k_{\ell}\,\p_{\bar\tau}\,\wt H_{\ell}\=\frac{i}{2\pi}\,\p^{2}_{\bar u}\, \wt H_{\ell}\,.
\ee
We have that the function~$\chi_\text{ell}(\wt M_\text{tor}; \tau,z)$ obeys:
\bea
&& \p_{\overline{\t}} \chi_\text{ell}(\wt M_\text{tor}; \tau,z)  \cr
&& \qquad  \= \bigintsss_{E_\tau^{d}} \, \prod_{\ell=1}^d   \frac{\dd^2 u'_\ell}{\t_{2}} \,  \sum_{i=1}^{d}\p_{\bar\tau}
\wt H_i (\tau, z, u'^{i}) 
\prod_{\ell=1,\,\ell\neq i}^d \wt H_\ell (\tau, z, u'_\ell)  \; \chi_\text{ell}(M_\text{tor};\tau,z, u') \, , \cr
&& \qquad \=  \frac{i}{2\pi}\bigintsss_{E_\tau^{d}} \, \prod_{\ell=1}^d   \frac{\dd^2 u'_\ell}{\t_{2}} \,  
\sum_{i=1}^{d}\frac{1}{k_{i}}\p^{2}_{\bar u'^{i}} \wt H_i (\tau, z, u'_i) \prod_{\ell=1,\,\ell\neq i}^d 
\wt H_\ell (\tau, z, u'_\ell)  \; \chi_\text{ell}(M_\text{tor};\tau,z, u') \, , \cr
&& \qquad \= \frac{i}{2\pi}\sum_{i=1}^{d}\bigintsss_{E_\tau^{(d-1)}} \, \prod_{\ell=1 \atop \ell\neq i}^d  
\Bigl(\frac{\dd^2 u'_\ell}{\t_{2}} \wt H_\ell (\tau, z, u'_\ell)\Bigr) \,  \times  \label{holanom}  \\ 
&& \qquad\qquad\qquad\qquad \qquad \qquad
\times\qquad \frac{1}{k_{i}} \sum_{j=1}^{d}\oint_{v_{j}=0} du_{i} \, \p_{\bar u^{i}}  
\wt H_i (\tau, z, u'_i) \, \chi_\text{ell}(M_\text{tor};\tau,z, u')  \, , \cr
&& \qquad \= -\sum_{i,j=1}^{d}\bigintsss_{E_\tau^{(d-1)}} \, \prod_{\ell=1,\atop \ell\neq i}^d 
\Bigl(\frac{\dd^2 u'_\ell}{\t_{2}} \wt H_\ell (\tau, z, u'_\ell)\Bigr) \,  
\underset{v_{j}(u'_{i})=0}{\text{Res}} \Bigl(\chi_\text{ell}(M_\text{tor};\tau,z, u') \Bigr) \times    \cr
&& \qquad\qquad\qquad \qquad\qquad\qquad\qquad \qquad\qquad\qquad\qquad \qquad
\times\frac{1}{k_{i}} \p_{\bar u^{i}}  \wt H_i (\tau, z, u'_i )\mid_{v_{j}(u'_{i})=0}\,. \nn
\eea
Here we have used the property \eqref{anomalyEqu} of the function~$\wt H_{\ell}$ to obtain the second equality. 
To obtain the third equality we have used the fact that $\chi_\text{ell}(M_\text{tor};\tau,z, u')$ is holomorphic in $u'$ 
which allows us to write the integrand as a total derivative. 
We then use Stokes's theorem to convert the integral over $u'_{i}$ to a contour integral, which is
then evaluated using Cauchy's residue formula. 

For the simplest case~$d=1$ the holomorphic anomaly equation reduces precisely to the one 
obeyed by mock Jacobi forms (see Appendix~\ref{sec:Jacobi}). In this case the right-hand side 
of~\eqref{holanom} can be identified as a contribution from the compensator field including its 
winding and momentum modes around the asymptotic cylinder~\cite{Murthy:2013mya}. 
For higher~$d$ we see a nested structure---the right-hand side of the holomorphic anomaly equation 
is governed not only by products of the functions~$\wt H_{\ell}$ and their derivatives but also by the 
residues of~$\chi_{\ell}(M_\text{tor})$ at the points~$v_{i}(u)=0$, which are themselves meromorphic
Jacobi forms. It would be interesting to give a more precise physical interpretation along the lines 
of~\cite{Ashok:2011cy, Ashok:2013kk, Giveon:2015raa}.

\subsection{Examples}

Now we illustrate all this with our usual examples. 

\subsection*{$\text{Squashed}\,\,\bf \IC/\IZ_{2}$, $(n,d)=(2,1)$}

We start with the squashed version of the~$\IC/\IZ_{2}$ theory discussed in Section~\ref{sec:ellgenexamples}.
The original unsquashed gauge theory has a~$U(1)$ gauge group with two chiral superfields $\Phi_1, \Phi_2$ with 
charges~$Q_1=-Q_2=1$. Now we gauge the~$U(1)$ flavor symmetry under which the chiral multiplets have 
charges~$F_1, F_2$, respectively. In this case the partition function is (with $b=F_1+F_2$):
\bea
&& \chi_\text{ell}(\wt{\IC/\IZ}_{2}; \tau,z) \\
&& \qquad \=\frac{k}{\tau_2}\,\int_{E_\tau} \dd^2u'\,\frac{\vth_1(\tau,-z+b u')}{\vth_1(\tau,  bu')} \sum_{m,w\in\mathbb{Z}}\,
e^{2\pi i b w z}e^{-\frac{\pi k}{\tau_2}(w\tau+m+ u'+\frac{bz}{k})(w\bar\tau+m+\bar{u'}+\frac{bz}{k})}\,. \nn
\eea
One can also unfold the integration over $E_{\tau}$ to the entire complex plane to obtain:
\bea
&& \chi_\text{ell}(\wt{\IC/\IZ}_{2}; \tau,z) \= \cr
&&\quad \= \frac{k}{\tau_2}\,\int_{E_\tau} \dd^2u'\,\sum_{m,w\in\mathbb{Z}}\,
\frac{\vth_1(\tau,(-z+b (u'+m+w\tau))}{\vth_1(\tau,  b(u'+m+w\tau))}
e^{-\frac{\pi k}{\tau_2}(w\tau+m+ u'+\frac{bz}{k})(w\bar\tau+m+\bar{u'}+\frac{bz}{k})}\,,\nn\\
&&\quad \= \frac{k}{\tau_2}\,\sum_{m,w\in\mathbb{Z}}\int_{E_\tau} \dd^2u'\,\,
\frac{\vth_1(\tau,(-z+b (u'+m+w\tau))}{\vth_1(\tau,  b(u'+m+w\tau))}
e^{-\frac{\pi k}{\tau_2}(w\tau+m+ u'+\frac{bz}{k})(w\bar\tau+m+\bar{u'}+\frac{bz}{k})}\,,\nn\\
&&\quad \= \frac{k}{\tau_2}\,\int_{\IC} \dd^2u'\,\frac{\vth_1(\tau,-z+b u')}{\vth_1(\tau,  bu')}
e^{-\frac{\pi k}{\tau_2}(u'+\frac{bz}{k})(\bar{u'}+\frac{bz}{k})} \, . 
\eea
Changing the integration variable to $\wt u=bu'$, we get (with~$\wt k=k/b^2$)
\be
\chi_\text{ell}(\wt{\IC/\IZ}_{2}; \tau,z)\=\frac{\wt k}{\tau_2}\,\int_{\IC} 
\dd^2 \wt u\,\frac{\vth_1(\tau,-z+\wt u)}{\vth_1(\tau,  \wt u)}
e^{-\frac{\pi \wt k}{\tau_2}(\wt u+\frac{z}{\wt k})(\bar{\wt u}+\frac{z}{\wt k})} \, .
\ee
This can also be written in terms of integral over single torus~$E_\t$ (by folding back on to the torus) as
\bea
&& \chi_\text{ell}(\wt{\IC/\IZ}_{2}; \tau,z) \cr
&&  \qquad \= \frac{\wt k}{\tau_2}\,\int_{E_\t} \dd^2 u\,
\sum_{m,w\in\IZ}\frac{\vth_1(\tau,-z+ u+m\t+w)}{\vth_1(\tau,   u+m\t+w)}
e^{-\frac{\pi \wt k}{\tau_2}(u+m\t+w+\frac{z}{\wt k})(\bar{ u}+m\bar\t+w+\frac{z}{\wt k})} \, ,\nn\\
&& \qquad \= \frac{\wt k}{\tau_2}\,\int_{E_\t} \dd^2 u\,\frac{\vth_1(\tau,-z+ u)}{\vth_1(\tau, u)}
\sum_{m,w\in\IZ}e^{2\pi imz}e^{-\frac{\pi \wt k}{\tau_2}( u+m\t+w+\frac{z}{\wt k})(\bar{ u}+m\bar\t+w+\frac{z}{\wt k})}\,.
\eea
This is precisely the elliptic genus of the cigar conformal field theory with 
central charge~$c=3(1+\frac{2}{\wt k})$ \cite{Troost:2010ud,Eguchi:2010cb,Ashok:2011cy}. 
The function~$\chi_\text{ell}(\wt{\IC/\IZ}_{2}; \tau,z)$ transforms like a Jacobi form of weight~$0$ and 
index~$m\=\frac{1}{2}+\frac{1}{\wt k}$,  
as consistent with the fact that the model flows to a superconformal field theory with central charge~$c=6m$. 
The function~$\chi_\text{ell}(\wt{\IC/\IZ}_{2}; \tau,z)$ obeys the holomorphic anomaly equation:
\be
\p_{\overline{\t}} \chi_\text{ell}(\wt M_\text{tor}; \tau,z)\=\frac{\wt k}{2\tau^{2}_{2}}\frac{\vth_{1}(\t,-z)}{\eta(\t)^{3}}
\sum_{m,w}e^{2\pi iwz}e^{-\frac{\pi k}{\tau_{2}}(w\t+m+\frac{z}{k})(w\bar\t+m+\frac{z}{k})}(w\t+m+\frac{z}{k})\,,
\ee
which is precisely the definition of a mixed mock Jacobi form whose shadow is a linear combination of 
products of weight~$3/2$ and weight~$1/2$ theta functions~\cite{Dabholkar:2012nd, Murthy:2013mya}.

\subsection*{Squashed $A_{1}$}

Next we consider the squashed version of the~$A_{1}$ space. We recall that the unsquashed model is a~$U(1)$ 
gauge theory with three chiral multiplets with charges~$1,-2,1$, respectively. The squashing gauges the~$U(1)^{2}$ flavor 
symmetry under which the chiral fields~$\Phi_{i}$, $i=1,2,3$, have charges~$F^{\ell}_{i}$, $\ell=1,2$. 
The elliptic genus of the squashed 
model is (with~$b_{\ell}=\sum_{i=1}^3F_{i}^{\ell}$):
\bea
\chi_\text{ell}(\wt A_{1};\tau,z)&=&\frac{k_1k_2}{\tau^2_2}\int_{E_\tau} \dd^2u_1'\int_{E_\tau}\,
\dd^2u_2'\,\chi_\text{ell}(A_{1};\tau,z,u')\times\,\nn\\
&& \qquad \times \sum_{m_{1,2},w_{1,2}\in\mathbb{Z}}e^{2\pi i (b_1w_1+b_2w_2)z}
e^{-\frac{\pi k_1}{\tau_2}(w_1\tau+m_1+u'_1+\frac{b_1z}{k_1})(w_1\bar\tau+m_1+\bar{u'_1}+\frac{b_1z}{k_1})} \times\cr
&& \qquad \qquad \qquad \qquad 
\times \; e^{-\frac{\pi k_2}{\tau_2}(w_2\tau+m_2+u'_2+\frac{b_2z}{k_2})(w_2\bar\tau+m_2+\bar{u'_2} +\frac{b_2z}{k_2})}\,.
\eea
The function~$\chi_\text{ell}(\wt A_{1}; \tau,z)$ transforms like a Jacobi form of weight $0$ and 
index $m=1+\frac{1}{\wt k_{1}}+\frac{1}{\wt k_{2}}$,
and obeys the holomorphic anomaly equation: 
\bea
\p_{\bar\t}\chi_\text{ell}(\wt A_{1}; \tau,z)&=& -\bigintsss_{E_\tau}   \frac{\dd^2 u'_1}{k_{2}\t^{2}_{2}} \,\wt H_1 (\tau, z, u'_1)  \underset{v_{j}(u'_{2})=0}{\text{Res}} \chi_\text{ell}(M_\text{tor};\tau,z, u') 
 \,\p_{\bar u'_{2}}  \wt H_i (\tau, z, u'_i )\mid_{v_{j}(u'_{2})=0} \nn\\
 &&-\bigintsss_{E_\tau}   \frac{\dd^2 u'_2}{k_{1}\t^{2}_{2}} \,\wt H_2 (\tau, z, u'_2)   
\underset{v_{j}(u'_{1})=0}{\text{Res}} \chi_\text{ell}(M_\text{tor};\tau,z, u') \,\p_{\bar u'_{1}}  \wt H_i (\tau, z, u'_i )\mid_{v_{j}(u'_{1})=0}\nn\,.\\
\eea
For simplicity we consider here the case for the following flavor charges:
\be
\begin{tabular}{ l | c | r }			
   & $U(1)_{1}$ &$U(1)_{2}$  \\
  \hline
  $\phi_{1}$ & 1 & 0 \\
  $\phi_{2}$ & 1 & 1 \\
$\phi_{3}$ & 0 & 1 \\ 
\end{tabular} \quad ,
\ee
so that the poles in $\chi_\text{ell}(A_{1};\tau,z,u')$ are 
at~$3u'_{1}+u'_{2}=0$, $3u'_{2}+u'_{1}=0$ and $u'_{1}-u'_{2}=0$. The holomorphic anomaly equation is:
\bea
&& \p_{\bar\t}\chi_\text{ell}(\wt A_{1}; \tau,z) \= \cr
&& \qquad \= -\frac{\vth_1(\t,-z)}{2\pi \eta(\t)^{3}k_{1}}\int\,\frac{\dd^{2}u'_{2}}{\t^{2}_{2}} \wt H_{2}
\sum_{a,b=0}^{2}\left[\p_{\bar u'_{1}} \wt H_{1}|_{u'_{1}=\frac{1}{3}(-u'_{2}+a+b\t)}\,
e^{2\pi ibz}\frac{\vth_1(\t,-z+\frac{4u'_{2}-a-b\t}{3})}{\vth_1(\t,\frac{4u'_{2}-a-b\t}{3})}\right]\nn\\
&& \qquad\qquad -\frac{\vth_1(\t,-z)}{2\pi \eta(\t)^{3}k_{1}}\int\,\frac{\dd^{2}u'_{2}}{\t^{2}_{2}} \wt H_{2}
\left[\p_{\bar u'_{1}} \wt H_{1}|_{u'_{1}=-3u'_{2}}\,\frac{\vth_1(\t,-z-4u'_{2})}{\vth_1(\t,-4u'_{2})}\right]\\
&& \qquad \qquad -\frac{\vth_1(\t,-z)}{2\pi \eta(\t)^{3}k_{2}}\int\,\frac{\dd^{2}u'_{1}}{\t^{2}_{2}} \wt H_{1}
\sum_{a,b=0}^{2}\left[\p_{\bar u'_{2}} \wt H_{2}|_{u'_{2}=\frac{1}{3}(-u'_{1}+a+b\t)}\,e^{2\pi ibz}
\frac{\vth_1(\t,-z+\frac{4u'_{1}-a-b\t}{3})}{\vth_1(\t,\frac{4u'_{1}-a-b\t}{3})}\right]\nn\\
&& \qquad \qquad-\frac{\vth_1(\t,-z)}{2\pi \eta(\t)^{3}k_{2}}\int\,\frac{\dd^{2}u'_{1}}{\t^{2}_{2}} \wt H_{1}
\left[\p_{\bar u'_{2}}\wt H_{2}|_{u'_{2}=-3u'_{1}}\,\frac{\vth_1(\t,-z-4u'_{1})}{\vth_1(\t,-4u'_{1})}\right]\nn\,.
\eea
The squashed model is conjectured to flow to an SCFT with central charge~$c=6m$ that arises 
(in the case~$k_{1}=k_{2}$) on a NS5-brane in string theory wrapped on~$\IC\IP^{1}$ \cite{Hori:2002cd}.

\vspace{0.2cm}

\subsection*{Squashed Conifold}
Our third example is squashed conifold. The unsquashed model has one~$U(1)$ gauge field and 
four chiral superfields with charges~$(+1,+1,-1,-1)$. There is a~$U(1)^{3}$ flavor symmetry 
under which the chiral superfields~$\Phi_{i}$ $i=1,..,4$, have charges~$F^{\ell}_{i}$, $\ell=1,2,3$. 
The elliptic genus in this case is:
\bea
\chi_\text{ell}(\wt{\text{Conifold}};\tau,z) 
&\=& k_1k_2 k_{3} \int_{E_\tau} \prod_{\ell=1}^3 \frac{\dd^2u'_\ell}{\tau_2} \, \chi_\text{ell}(\text{Conifold};\tau,z,u')\times\\
&&\sum_{m_\ell,w_\ell \in\mathbb{Z}}\,e^{2\pi i\sum_{\ell=1}^3b_\ell w_\ell z}\, 
e^{-\sum_{\ell=1}^3\frac{\pi k_\ell}{\tau_2}(w_\ell\tau+m_\ell+u'_\ell+\frac{b_\ell z}{k_\ell})(w_\ell\bar\tau
+m_\ell+\bar{u'_\ell}+\frac{b_\ell z}{k_\ell})}\,.\nn
\eea
The function~$\chi_\text{ell}(\wt{\text{Conifold}};\tau,z)$ transforms like a Jacobi form of weight $0$ and 
index $m=(\frac{3}{2}+\sum_{\ell=1}^{3}\frac{1}{\wt k_{\ell}})$, and obeys a holomorphic anomaly equation 
as above. The squashed model is conjectured to flow to an SCFT with central charge~$c=6m$ that arises 
(in the case that all the~$k_{i}$ are equal) on a NS5-brane in string theory wrapped 
on~$\IC\IP^{2}$ \cite{Hori:2002cd}.

\section{A compact example \label{sec:Compact}}

In this section we study the simplest example of a squashed toric sigma model with compact target space,
namely the supersymmetric sausage discussed in Section \ref{sec:Squashed}. This model has a mass gap 
and is therefore expected to flow to a trivial theory in the IR. The elliptic genus is
not well-defined because, as explained in Section \ref{sec:EllGenTor}, the continuous~$R$-symmetry 
is not conserved. This is also manifested in the computation because under the elliptic 
transformation~$u\rightarrow u+\l\t+\m$, with~$\l,\mu \in \IZ$, 
the one loop determinant for~$\CP^{1}$ picks up a phase~$e^{4\pi iz\l}$. 
However, from this we see that it is well defined for discrete values of~$z=0$ and~$z=\frac{1}{2}$, 
corresponding to the discrete $R$-symmetry, for which we have the Witten index ($z=0$) and 
twisted Witten index ($z=\frac{1}{2}$). The twisted Witten index for~$\CP^{1}$ is zero, as can be 
seen from the formula \eqref{JKresidue1}, and thus, using the formula \eqref{SquashElliptic3}, 
it is also zero for squashed~$\CP^{1}$. Therefore here we will only discuss the Witten index which 
equals~2 for~$\CP^{1}$. The computation of the Witten index of the squashed model 
is slightly subtle. The result, as we discuss below, is that it is also independent of the 
squashing deformation. 

The GLSM describing squashed~$\IC\IP^{1}$ has two chiral multiplets with gauge~$U(1)$ charges $(+1,+1)$ 
and with~$U(1)$ flavor charges~$F_{i}$. 
One method to compute the Witten index of the squashed theory is to use the same techniques as those of 
Section~\ref{sec:EllGen} with $z=0$. The starting point of this method would be the elliptic genus of~$\CP^{1}$ 
in the form of Equation~\eqref{JKresidue1} evaluated at~$z=0$. This quantity, however, is divergent and 
needs to be regulated. We use the trick of~\cite{Benini:2013nda} (in a slightly modified version suitable 
for our purposes) and introduce an extra chiral superfield of gauge charge $-2$. The elliptic genus of this 
model is now well defined as the sum of the gauge charges is zero. We can then squash this model with 
respect to the original flavor symmetries, compute the elliptic genus of this model as in the previous section,
and then set~$z=0$. At the end one introduces a twisted mass for the extra chiral multiplet by giving a vev 
to the scalar field  (this twisted mass breaks the chiral R-symmetries, but one can consistently turn it on at~$z=0$) 
in the extra flavor symmetry background vector multiplet. This decouples the extra chiral multiplet so that we 
get the vacuum manifold of~$\wt{\IC\IP^{1}}$.

In practice, one notices that the extra chiral multiplet does not modify the location of poles if we choose 
to evaluate the residues at~$\mathfrak{M}_\text{sing}^+$. Furthermore, the extra chiral multiplet
also does not contribute to the one-loop determinant at~$z=0$, and so we can essentially ignore it. 
These considerations lead to the following expression for the elliptic genus for the deformed model (we will continue to denote it as ~$\wt{\IC\IP^{1}}$):
\bea
&& \chi_\text{ell}(\wt{\IC\IP^{1}};\tau,z) \\
&& \qquad  \=\frac{ik}{\tau_2}\,\int_{E_\tau} \dd^2u'\,\frac{\eta(\tau)^3}{\vth_1(\tau,z)}\sum_{m,n \in \IZ} 
\sum_{u_j\in \mathfrak{M}_{\text{sing}}^+}\oint_{u=u_j} du\, Z^{(m,n)}_{\text{chiral}} \, e^{-\frac{\pi k}{\tau_2}(m\tau+n+u'+\frac{b_{1}z}{k})(m\bar\tau+n+\bar{u'}+\frac{b_{1}z}{k})} \, , \nn
\eea
where $Z^{(r,s)}_{\text{chiral}}(\tau,z,u,u')$ is independent of~$s$ and is given by
\be
Z^{(r,s)}_{\text{chiral}}(\tau,z,u,u')\=e^{2\pi i (F_{1}-F_{2}) rz} \, 
\frac{\vth_1(\tau,-z+u+ F_1u')}{\vth_1(\tau,u+ F_1u')}\frac{\vth_1(\tau,-z+u+ F_2u')}{\vth_1(\tau,u+ F_2u')} \, . 
\ee
Performing the contour integral and picking up the poles at $u=-F_{1}u'$ and $u=-F_{2}u'$, we obtain
\bea
&& \chi_\text{ell}(\wt{\IC\IP^{1}};\tau,z)\=\frac{k}{\tau_2}\,\int_{E_\tau} \dd^2u'
\left[\frac{\vth_1(\tau,-z+(F_1-F_2)u')}{\vth_1(\tau,(F_1-F_2)u')}+
\frac{\vth_1(\tau,-z+(F_2- F_1)u')}{\vth_1(\tau,(F_2-F_1)u')}\right]\times\nn\\
&& \qquad \qquad  \qquad \qquad \qquad \qquad \times \sum_{m,n\in \IZ}e^{2\pi i (F_{1}-F_{2})mz} 
e^{-\frac{\pi k}{\tau_2}(m\tau+n+u'+\frac{b_{1}z}{k})(m\bar\tau+n+\bar{u'}+\frac{b_{1}z}{k})}\,.
\eea

We now substitute $z=0$ to obtain the Witten index of~$\wt{\IC\IP^{1}}$:
\bea
\chi(\wt{\IC\IP^{1}};\tau)\= \chi_\text{ell}(\wt{\IC\IP^{1}};\tau,z= 0) \=\frac{2k}{\tau_2}\,\sum_{m,n\in \IZ} 
\int_{E_\tau} \dd^2u' \,\,e^{-\frac{\pi k}{\tau_2}(m\tau+n+u')(m\bar\tau+n+\bar{u'})}\,.
\eea
We can evaluate the above integral in two ways. The first way is to unfold the domain of the integration 
over~$E_\tau$ to the entire complex plane to obtain
\be
\chi(\wt{\IC\IP^{1}};\tau)
\=\frac{2k}{\tau_2}\sum_{m,n\in \IZ} \int_{E_\tau} \dd^2u' \,\,e^{-\frac{\pi k}{\tau_2}(m\tau+n+u')(m\bar\tau+n+\bar{u'})} 
\= \frac{2k}{\tau_2}\int_{\IC}\dd^2u' \,e^{-\frac{\pi k}{\tau_2}|u'|^2} \=2\,.
\ee
The second way is to keep the domain of the integration fixed and perform the sum over the integrand 
using Poisson resummation formula which gives (with~$u=u_{1}+i u_{2}$):
\bea
\chi(\wt{\IC\IP^{1}};\tau)
&\=&\frac{2}{\tau_2}\,\sum_{p,r\in \IZ} 
\int_{E_\tau} \dd^2u' \,\,e^{-\frac{\pi}{k\tau_2}(p-r\tau_2)^2}e^{\frac{2\pi i}{\tau_2}[(p-r\tau_1) u'_{2} 
+r \tau_2 u'_{1}]}\,,\nn\\
&\=&2\,\sum_{p,r\in \IZ}e^{-\frac{\pi}{k\tau_2}|p-r\tau_2|^2}\delta_{p,0}\delta_{k,0}\=2\,.
\eea
Thus we see that the Witten index of squashed ${\IC\IP^{1}}$ is $2$, i.e.~it is independent of the 
squashing deformation.

\vspace{1cm}

\section*{Acknowledgements}
We thank Amihay Hanany, Sungjay Lee, Dario Martelli, Dimitri Panov, Cristian Vergu, 
and Katrin Wendland for useful conversations. 
R.~G.~would like to thank the ASICTP for hospitality during the final stages of this work.
This work was supported by the EPSRC First Grant UK EP/M018903/1 and and by 
the ERC Consolidator Grant N.~681908, ``Quantum black holes: A macroscopic window into the 
microstructure of gravity''.

\vspace{0.4cm}

\appendix

\section{Multi-variable Jacobi forms and mock Jacobi forms \label{sec:Jacobi}}

The classical theory of (one-elliptic-variable) Jacobi forms (see e.g.~\cite{Eichler:1985ja}) deals with 
a holomorphic function~$\v(\tau, u)$ from~$\mathbb{H} \times\IC$ to $\IC$ which is ``modular in $\tau$ and elliptic in $u $'' 
in the sense that it transforms under the modular group as 
  \be\label{modtransform}  \v\Bigl(\frac{a\t+b}{c\t+d},\frac{z}{c\t+d}\Bigr) \= 
   (c\t+d)^k\,e^{\frac{2\pi imc z^2}{c\t+d}}\,\v(\t,z)  \qquad \forall \quad
   \Bigl(\begin{array}{cc} a&b\\ c&d \end{array} \Bigr) \in SL(2; \IZ) \ee
and under the translations of $z$ by $\mathbb{Z} \tau + \mathbb{Z}$ as
  \be\label{elliptic}  \v(\t, z+\lambda\tau+\mu)\= e^{-2\pi i m(\lambda^2 \t + 2 \lambda z)} \v(\t, z)
  \qquad \forall \quad \lambda,\,\mu \in \IZ \, . \ee
The number~$k\in \half \IZ$ is called the weight and~$m\in \half \IZ$ is called the index of the Jacobi form.

In the text we also deal with Jacobi forms of~$n$ elliptic variables~$\vec{z}=(z_{1},\cdots, z_{n})$ and one modular variable~$\t$,
which are meromorphic functions of~$z_{i} \in \IC$ and~$\tau \in \IH$. Now the index~$M$ becomes matrix-valued 
with entries~$M_{ij}$, $i,j=1,\cdots n$. The main transformation properties~\eqref{modtransform}
and~\eqref{elliptic} now become: 
  \be\label{modtransform}  \v\Bigl(\frac{a\t+b}{c\t+d},\frac{\vec{z}}{c\t+d}\Bigr) \= 
   (c\t+d)^k\,
  \exp \Bigl(2 \pi i \frac{c}{c\t+d} \vec{z}^\text{t} M \vec{z}  \Bigr)  \,\v(\t,\vec{z}) \, ,   \ee
and, with~$\vec{\lambda} = (\l_{1}, \cdots, \l_{n})$, $\vec{\mu} = (\mu_{1}, \cdots, \mu_{n})$,
  \be\label{elliptic}  \v(\t, \vec{z} + \vec{\lambda} \, \tau+\vec{\mu})\= 
 \exp\Bigl(- 2 \pi i \bigl( \vec{\lambda}^\text{t} M \vec{\lambda} \, \t + 2 \vec{\lambda}^\text{t} M \vec{z} \bigr) \Bigr)  \v(\t, \vec{z})
  \qquad \forall \quad \lambda_{i},\,\mu_{i} \in \IZ \, . \ee

\subsection*{Some modular and Jacobi forms}

Some functions that appear in the equations in this paper are the Dedekind eta function, a modular 
form of weight~$1/2$:
\be\label{defeta}
  \eta(\tau)\;:=\; q^{1/24} \prod_{n=1}^{\infty} (1-q^{n}) \, , 
\ee
and the odd Jacobi theta function which is a Jacobi form of weight~$1/2$ and index~$1/2$:
\be
\vartheta_1(\tau,z) = -iq^{1/8} \zeta^{1/2} \prod_{n=1}^\infty (1-q^n) (1-\zeta q^n) (1-\zeta^{-1} q^{n-1}) 
= i \sum_{m \in \IZ} e^{\pi i (m+\half)}  \, q^{(m+1/2)^2/2} \, \zeta^{m+\half} \, . 
\ee
These two functions obey the relation:
\be\label{etatheta} 
 \left. \frac{1}{2\pi i} \, \frac{d}{dz} \vth_{1}(\t,z) \right|_{z=0} \= -i \, \eta(\t)^{3} \, . 
\ee

We use below, for $\,\ell\in\IZ/2m\IZ$, the standard theta function 
\be\label{deftheta1/2} 
  \vth_{m,\ell}(\t,z) \=  \sum_{\l\inn\IZ \atop \l = \ell \mypmod{2m}} q^{\l^2/4m} \, \zeta^{\l},  
\ee
and its first Taylor coefficient 
\be\label{deftheta3/2} 
  \vth^{(1)}_{m,\ell}(\t) \=  \left. \frac{1}{2\pi i} \frac{d}{dz} \vth_{m,\ell}(\t,z) \right|_{z=0} \= 
  \sum_{\l\inn\IZ \atop \l = \ell \mypmod{2m}}\l\,q^{\l^2/4m} \, .
\ee

\vspace{0.2cm}

\subsection*{The cigar elliptic genus}

In the main text the modular properties and, in particular, the holomorphic anomaly equation obeyed 
by the cigar elliptic genus was referred to a few times. Here we summarize a few of the important 
formulas taken from~\cite{Murthy:2013mya}. 
The elliptic genus of the cigar coset at level~$k$ transforms as a holomorphic Jacobi form of weight~$0$
and index~$\frac12 + \frac1k$, and obeys the holomorphic anomaly equation:
\be \label{cigfreebos}
\p_{\bar{\tau}} \chi^{\rm cig}(\t,z)  =  -\frac{k}{4 \pi\t_{2}} \, \frac{\vth_1(\t,z) }{\eta(\t)^{3}} \, 
\left. \p_{\bar{u}} \sum_{m,w \in \IZ} \,  e^{2 \pi i z w - \frac{\pi k}{\t_{2}} |m + w \t + u +\frac{z}{k} |^{2}} \right|_{u=0} \, ,
\ee
or, equivalently, using Poisson resummation:
\be \label{L0barexpect}
\p_{\bar{\tau}} \chi^{\rm cig}(\t,z) =  \frac{i\sqrt{k}}{2\sqrt{\t_{2}}} \, \frac{\vth_1(\t,z) }{\eta(\t)^{3}} \, 
\sum_{n,w \in \IZ} \, (n-wk) \, q^{\frac{(n+wk)^{2}}{4k}} \, \overline{q}^{\frac{(n-wk)^{2}}{4k}}  \,  \zeta^{-\frac{n}{k}+w}  \, . 
\ee
The right-hand side of the above equation can be written in terms of 
standard~$\vth_{m,\ell}$ functions to obtain:
\be \label{holanomchi}
-\frac{2i}{\sqrt{k}} \t_{2}^{1/2} \p_{\bar{\t}} \chi^{\rm cig}(\t,z)  = \frac{1}{k} \frac{\vth_1(\t,z) }{\eta(\t)^{3}} \, 
 \sum_{\a,\b\in \IZ/2k\IZ}  \, e^{2\pi i \frac{\a \b}{k}} \, q^{\frac{\a^{2}}{k}} \, \zeta^{\frac{2\a}{k}} \, 
 \sum_{\ell \; (\text{mod} \, {2k})}  \, \overline{\vth_{k,\ell}^{(1)} (\t)} \; \vth_{k,\ell}\big(\t,\frac{z+\a\t+\b}{k}\big) \, ,
\ee
from which we see that it is a mixed mock Jacobi form whose shadow is given by the right-hand side of this 
equation.

\section{Holomorphic construction of toric manifolds \label{holtor}}

Toric manifolds (or toric varieties) can be thought of as a generalization of complex projective spaces~$\IC\IP^{n}$, 
which we assume are reasonably familiar to the reader. We recall that~$\IC\IP^{n} = (\IC^{n+1} - \{0\})/\IC^{*}$,
where an element~$\lambda \in \IC^{*}$ acts on the coordinates~$z_{i}$, $i=1,\cdots,n+1$, of the~$\IC^{n+1}$ 
as~$z_{i} \mapsto \lambda z_{i}$.  
A general complex~$d$-dimensional toric variety is the quotient space\footnote{We are suppressing here the 
additional possibility of discrete quotients.}
\be
V \= (Y-F)/T \, , 
\ee
where~$V = \IC^{n}$, $F \subset V$ 
is a union of hyperplanes passing through the origin, and the torus~$T = (\IC^{*})^{n-d}$.
An element~$g_{a} \in T$, $a\=1,\cdots, n-d$, acts on the coordinates~$z_{i}\in V$, $i=1,\cdots, n$ as:
\be \label{u1action}
g_{a}(\l): z_{i} \mapsto \l^{Q^{a}_{i}} z_{i} \, , \quad \text{for some~$\l \in \IC^{*}$.}
\ee
The charges~$Q^{a}_{i}$ as well as the precise description of~$F$ are determined by combinatorial 
data, together called a \emph{fan}, which completely determine~$V$. This construction of a toric manifold is called 
the \emph{holomorphic quotient construction}.

On restricting~$|\l|=1$ in~\eqref{u1action}, we obtain the action of the real torus~$U(1)^{d}$ on the manifold and,
in particular, the points where this torus action has fixed points. This allows us to represent the toric manifold 
in terms of a so-called \emph{toric diagram}. The simplest example is the 
one-dimensional case~$\IC\IP^{1}$ which is defined by pairs~$(z_{1}, z_{2}) \in \IC^{2}$ with the 
identification~$(z_{1},z_{2}) \sim \l (z_{1},z_{2})$, $\l \in \IC$. When~$z_{1}\neq 0$, 
we have~$(z_{1},z_{2}) \sim (1,z)$
with~$z \in \IC$. The~$U(1)$ acts as~$z\mapsto e^{i\v} z$, with~$\v \in \IR/\IZ$. The fixed points are at~$z=0$,
and~$w=1/z=0$. The~$\IC\IP^{1}$ can thus be drawn as a line segment with ends~$z=0$ and~$w=0$, with 
a circle fibered over it.

\section{Metric of two-dimensional toric manifolds \label{sec:2dtormetric}}

Consider a GLSM whose vacuum manifold is two-dimensional.
Let~$\rho$ be the magnitude of 
one of the chiral fields in the theory, in terms of which the symplectic form is written as:
\be
\o \=  \, \rho \, d\rho \wedge d\psi \, .
\ee
In order to perform the quotient construction described in Section~\ref{sec:GLSM} 
we consider the gauge-invariant (composite) field~$Z=R \, e^{i \psi}$. 
We solve the (algebraic) D-term constraints to get a function~$R(\rho)$,
and we denote the inverse function as~$\rho(R)$ (this may or may not be possible to do explicitly). We denote 
\be
' \, \equiv \, \frac{d}{d\rho} \,,  \qquad \dot{} \, \equiv \, \frac{d}{dR} \, . 
\ee
The symplectic form written in terms of the gauge-invariant variable is:
\be
\o \= \, \rho \, \dot{\rho}  \, dR \wedge d\psi  \=\frac{i}{2} \frac{\rho \, \dot{\rho}}{R} \, dZ \wedge d \overline{Z} \, ,
\ee
from which we can write the K\"ahler metric 
\be
ds^{2} \= \frac{\rho(R) \, \dot{\rho}(R)}{R} \, \bigl( dR^{2} + R^{2} d\psi^{2} \bigr) \, .
\ee
This formula is to be thought of as a function of the gauge-invariant variable~$Z$ or, equivalently, its magnitude~$R$ and angle~$\psi$.
We can also write this metric in terms of the coordinate~$\rho$ as:
\be
ds^{2} \= \rho \, \Bigl( \frac{R'(\rho)}{R(\rho)} \, d\rho^{2} + \frac{R(\rho)}{R'(\rho)} \, d\psi^{2} \Bigr) \, .
\ee

\section{GLSM action}\label{glsmLagr}

In the Lorentizian space, the Lagrangian of~$\mathcal N=(2,2)$ GLSM in the component form is given by
\bea
\mathcal L&= & \sum_{i=1}^n\Big[-D^\mu\bar\phi_i D_\mu\phi_i+i\bar\psi_{i-}(D_0+D_1)\psi_{i-}+i\bar\psi_{i+}(D_0-D_1)\psi_{i+}+\hat D_i|\phi_i|^2+|F_i|^2\nn\\
&&-|\hat\sigma_i|^2|\phi_i|^2-\bar\psi_{i-}\hat\sigma_i\psi_{i+}-\bar\psi_{i+}\hat{\bar\sigma}_i\psi_{i-}-i\bar\phi_i\hat\lambda_{i-}\psi_{i+}+i\bar\phi_i\hat\lambda_{i+}\psi_{i-}+i\bar\psi_{i+}\hat{\bar\lambda}_{i-}\phi_i\nn\\
&&-i\bar\psi_{i-}\hat{\bar\lambda}_{i+}\phi_i\Big]+\sum_{\ell=1}^d\frac{k_\ell}{2}\Big(-D^{'\mu}\bar p_\ell D'_\mu p_\ell+i\bar\chi_{\ell-}(\p_0+\p_1)\chi_{\ell-}+i\bar\chi_{\ell+}(\p_0-\p_1)\chi_{\ell+}\nn\\
&&+D'_\ell(p_\ell+\bar p_\ell)+|F_{p_\ell}|^2-|\sigma'_\ell|^2+i\chi_{\ell+}\lambda'_{\ell-}-i\chi_{\ell-}\lambda'_{\ell-}+i\bar\chi_{\ell+}{\bar\lambda'}_{\ell+}-i\bar\chi_{\ell-}{\bar\lambda}'_{\ell+}\Big)\nn\\
&&+\sum_{a=1}^{n-d}\frac{1}{2e_a^2}\Big(-\p^\mu{\bar\sigma_a}\p_\mu\sigma_a+i{\bar\lambda}_{a-}(\p_0+\p_1)\lambda_{a-}+i{\bar\lambda}_{a+}(\p_0-\p_1)\lambda_{a+}+ F_{a\, 01}^{2}+ D_a^2\Big)\nn\\
&&+\sum_{\ell=1}^d\frac{1}{2\wt e_\ell^2}\Big(-\p^\mu{\bar\sigma'_\ell}\p_\mu\sigma'_\ell+i{\bar\lambda'}_{\ell-}(\p_0+\p_1)\lambda'_{\ell-}+i{\bar\lambda'}_{\ell+}(\p_0-\p_1)\lambda_{\ell+}+F_{\ell\,01}^{'2}+ D_\ell^{'2}\Big)\Big]\nn\\&&-\sum_{a=1}^{n-d}r_aD_a\,.
\eea
Here $D_\mu$ is a covariant derivative with the combination of gauge fields $(Q^a_i V^a_m+F^l_i V^{l'}_m)$ and $D'_\mu$ is covariant derivative with gauge field $V^{l'}_m$. Other various fields are
\bea
\widehat D_i\= Q_i\cdot D+F_i\cdot D',\,\, \widehat\sigma_i\=\ Q_i\cdot \sigma+F_i\cdot \sigma',\,\,\widehat\lambda_i\=Q_i\cdot \lambda+F_i\cdot \lambda'
\eea
In the above expressions, we are using the following notations
\be
Q_i\cdot\sigma\=\sum_{a=1}^{n-d}Q_i^a\sigma^a,\quad \text{and} \quad F_i\cdot\sigma'\=\sum_{\ell=1}^dF_i^\ell\sigma'_\ell\,.
\ee
In going from Lorentzian to Euclidean space we replace~$x^{0}$ by~$ix^{2}$ and~$D_{a,\ell}$ by~$iD_{a,\ell}$.

%

\providecommand{\href}[2]{#2}\begingroup\raggedright\endgroup

\end{document}